\documentclass[twocolumn]{aastex62}

\usepackage{graphicx}

\graphicspath{{./}{figures/}}

\accepted{in Astrophysical Journal}

\shortauthors{Raghuvanshi et al.}

\begin{document}

\title{Simulating the collapse of rotating primordial gas clouds to study the survival possibility of Pop III protostars}
\author{Shubham P. Raghuvanshi}
\affil{Harish-Chandra Research Institute (HRI), Chhatnag Road, Jhusi, Prayagraj, 211019, Uttar Pradesh, India}
\author[0000-0002-6903-6832]{Jayanta Dutta}
\affil{Harish-Chandra Research Institute (HRI), Chhatnag Road, Jhusi, Prayagraj, 211019, Uttar Pradesh, India}

\begin{abstract}
It has been argued that the low-mass primordial stars ($m_{\rm Pop III}\,\leq 0.8\,M_\odot$) are likely to enter the main sequence and hence possibly be found in the present-day Galaxy. However, due to limitations in existing numerical capabilities, current three-dimensional (3D) simulations of disk fragmentation are capable of following only a few thousands of years of evolution after the formation of the first protostar. In this work we use a modified version of {\sc Gadget}-2 smoothed particle hydrodynamics(SPH) code to present the results of non-linear collapse of the gas clouds associated with various degrees of initial solid body rotation (parameterized by $\beta$) using a piecewise polytropic equation of state. The 3D simulations are followed till the epoch when 50$M_{\odot}$ of mass has been accreted in protostellar objects, which is adequate enough to investigate the dynamics of the protostars with the surrounding gaseous medium and to determine the mass function, accretion rate and survival possibility of these protostellar objects till present epoch. We found that evolving protostars that stay within slow-rotating parent clouds can become massive enough due to accretion in the absence of radiative feedback, whereas $10-12 \%$ of those formed within a fast-rotating clouds ($\beta \ge 0.1$) have the possibility to get ejected from the gravitational bound cluster as low mass stars.
\end{abstract}


\section{Introduction} 
\label{sec:intro} 
The age of the Universe and the expected time at which the very first stars formed makes direct observations a difficult prospect \citep[see recent survey, e.g.,][]{frebel19,sdb20,suda21,hartwig22, finkelstein22}. Theoretical prediction of $\Lambda$CDM model show that the entire process is led by gravitational collapse of dark matter halos as a consequence of hierarchical structure formation \cite[see the latest results, e.g.,][]{springel20,wang20,bohr20,may21,latif22}. At the time of collapse the primordial gas in halos is very hot and remains spread out due to its high pressure \citep{barrow17,chon18,barkana18}. Gas cools by radiating away energy and collapse to form a thin rotating circumstellar disk that grows over time and fragments due to gravitational and spiral-arm instability \citep{iy20,wollenberg20,chiaki22}. Some of the fragments that go on to become stars are not isolated and continue to interact with the surrounding gas. This interaction leads to an increase in mass of fragments as well as changes in their orbits. This leads to a very basic question of what is the final fate of these evolving fragments in the cluster. Do they merge with the central star \citep{kulkarni19,klessen19}, or do they move away from the cluster after their dynamical interaction with each other and with the surrounding gas \citep{sharda19,sugimura20}?  

It may happen that a fraction of them can either become massive due to rapid accretion \citep{umeda16,woods17,fukushima20} such that the resulting stars explode as (pair-instability) supernovae \citep{whalen14,welsh19,jeon21} or collapse to blackholes \citep{mr01,matsumoto15} or may lead to form supermassive blackhole \cite[SMBH:][]{alister20,herrington22}. There might also exist a fraction that remain as low mass protostars and hence can survive to the present day provided their mass remain as low as 0.8 $M_{\odot}$ \citep{marigo01,ishiyama16,susa19,dutta20}. Thus, the mass function of these fragments remains unclear and needs more investigation \cite[see review by][]{haemmerle20}. While it is possible to run detailed simulations of a few systems, it is difficult to explore a wide range of parameter space with this approach. When the numerical integration over density regime is computed beyond the formation of the first protostellar core, the collapse tends to be chaotic and highly nonlinear and becomes difficult to follow the dynamical system for long a time. As a consequence, current simulations lack the ability of following the evolution of fragments over a sufficient number of orbital revolutions within the disk.

In this paper, we aim to develop a model, building upon our earlier work \citep{dutta16} on the fragmentation of the unstable disk centred within the rotating collapsing gas clouds, using {\it modified version} of the Gadget-2 SPH simulations and piecewise polytropic equation of state, in order to put some upper bound on the final mass of the protostars after long time evolution of the gas. In the next section \S \ref{sec:nm} we describe in detail the initial conditions, implementation of polytropic index profile in the mathematical model and the modified numerical scheme. The details of dynamics are outlined in \S \ref{sec:result} with an emphasis on fragments that stay below the critical mass for surviving until present-day. We summarise work in \S \ref{sec:summary}.

\section{Numerical Methodology}  
\label{sec:nm}  
We start our discussion by considering uniform density spheres of gas with number density $n = 10^4 \text{cm}^{-3}$ and temperature $T = 250$ K, with initial solid body rotation. The gas density is represented by SPH particles. The clouds are numerically designed to model the local thermodynamic equilibrium (LTE) conditions for primordial gas and to study the effects of rotation on the time scales associated with the collapse and subsequent fragmentation. 

\subsection{Initial condition} 
\label{subsec:ic}
The gas clouds are modelled with approximately 2 million SPH particles, each with mass $m_{\rm gas}= 4.639 \times 10^{-4}$ $M_{\odot}$, uniformly distributed inside a sphere of radius equal to the Jeans radius at LTE i.e. $R$ = $R_{\rm J}$ $\approx$ 0.857 pc with total mass $M$ = $M_{\rm J}$ $\approx$ 940$M_{\odot}$. This numerical set-up allows us to follow the collapse accurately for about 10 orders of magnitude in density up to the number density $5 \times 10^{14} \text{ cm}^{-3}$ till the formation of first central sink (i.e., central hydrostatic core), and about 4 orders of magnitude in size up to about 10 AU. The mass resolution for $N_{\rm ngb}=100$ SPH neighbours is about 0.04639$M_{\odot}$. This implies that the rotating gas clouds are numerically well-resolved up to a critical number density $n_{\rm crit} = 2.02 \times 10^{15} \text{ cm}^{-3} $, given by 
\begin{equation}
n_{crit} = \left(\frac{3}{ 4\pi}\right) \left(  \frac{5K_BT}{G} \right)^3 \left(\frac{1}{\mu m_p}\right)^4 \left( \frac{1}{m_{\rm gas} N_{\rm ngb}} \right)^2\,,
\label{eqn_ncrit}
\end{equation} 
for temperature $T=1300$ K. Here $\mu$ is the hydrogen mass fraction of the gas, $m_{\rm p}$ is the mass of the proton and all other symbols have usual meaning. The free fall time over sound crossing time $t_{\rm ff}/t_{\rm sc} \sim 0.1$ for the clouds confirms the validity of the initial conditions for some degree of gravitational collapse.

Initial velocities are assigned to the SPH particles depending upon the angular velocity $(\Omega)$ of the clouds in addition to the thermal distribution of velocities and no internal turbulent motion. In absence of internal turbulent motions, the degree of rotation (i.e., strength of centrifugal support) of clouds is modelled by estimating the rotational energy over the total gravitational potential energy \cite[quantified by the parameter $\beta = R^3 \Omega^2/3GM$, as depicted in][]{sdz03}. We also model the distribution of angular momentum that originates either from distortion of the clouds or from their nonaxisymmetric nature due to differential rotation between the high and low- density regimes \citep{larson84,meynet02}.  
The gravitational forces from the dark matter are negligible compared to the self gravity of the gas on the length scales of our simulation, therefore for the sake of simplicity, we do not consider dark matter or the expansion of space itself in our simulations.

Although our calculation is based on the initial condition that assumes a spherical cloud with uniform density distribution, it is to be noted that molecular clouds are in general very irregular in shape. This is because in reality the formation of molecular clouds due to continual accumulation of baryonic matter within a dark matter halo encounters different types of forces occurring simultaneously \citep{larson72, shu77,bodenheimer81}.  For example, the interplay between the self-gravity of the cloud and the internal pressure of the infalling gas can introduce asymmetries or inhomogeneities in the clouds, which will be amplified during the collapse at a later time. That is why we see in the cosmological simulation that the initial density distribution of molecular clouds before it reaches LTE are non-homogeneities in nature and rather closer to the non-singular isothermal sphere \citep{suto88, on98}. 

In addition, as the gas cloud possesses small angular momentum, it soon experiences a differential rotation between the layers of the gas due to which the cloud becomes slightly centrally condensed. As a consequence, non-axisymmetric features are likely to appear at a very initial phase of formation of molecular clouds and even tend to wind up to form trailing spiral patterns at later stages of collapse. Larson 1984 \cite{larson84} has elaborated the fact that these tiny irregularities during the formation of the molecular cloud can certainly develop non-radial gravitational forces in a non-axisymmetric mass distribution. The associated gravitational/tidal torques that transfer angular momentum outward on an orbital time-scale is another reason for the non-homogeneities, which we see in cosmological simulations. Besides turbulent motion viscosity, and even magnetic fields can also play a role in shaping the clouds during its initial dynamical phase \citep{Truelove98, McKee02, Krumholz05}.

\subsection{Modeling Polytropic equation of state} 
\label{subsec:polyt}

In order to account for various heating and cooling processes happening simultaneously, which become important at certain number densities during the collapse, we model the thermal behaviour for the gas clouds \cite[following the discussion in][]{jappsen05} with a piecewise polytropic equation of state  
\begin{equation}
    T_i(n) = a_i n^{\gamma_i -1 } \text{                         ,                               }  i = 1,2,3\,.
\end{equation}
Here the polytropic index $\gamma_i$ changes values in a piecewise constant manner as a function of the number density $n$. The constant of proportionality $a_i$, which is initially determined from the thermal conditions of the gas is also rescaled in order to maintain the continuity of temperature across the certain intermediate densities $n_{\rm int}$
\begin{equation}
T_i(n_{\rm int}) = T_{i+1}(n_{\rm int})  
\end{equation}
according to the following equation 
\begin{equation}
a_{i+i} = a_i n_{\rm int}^{ \gamma_i -  \gamma_{i+1} } 
\label{eqn_gamma}
\end{equation}
The intermediate densities and the values for the polytropic index in different density intervals are chosen carefully in order to reproduce the temperature-density profile resulting from the primordial chemistry \citep{dnck15,pallottini17,bsg19}. Furthermore as the fast rotating clouds have larger time scales associated with the collapse and tend to have lower rates of compressional heating, they are significantly colder than their slow rotating counterparts \citep{dutta16}. Therefore all the values of the polytropic index also depend on the degree of rotation of the clouds. Table~1 summarizes the chosen values for the polytropic indices for all the clouds, where the piecewise polytropic index profile is divided in to three regions separated by intermediate densities $n_{\rm int} = 10^9, 10^{12}  \text{cm}^{-3} $. 

In order to implement general polytropic process in the publicly available {\sc Gadget}-2, we have added a polytropic index variable that controls the rate of change of entropic function same as the adiabatic index in original code. In addition to this, we have identified the original adiabatic index variable in the code with the quantity $1 + 1/C_V$, where $C_V$ is specific heat at constant volume for the gas. The implementation is explained in detail in the appendix section. This is necessary in order to accurately model the thermal and chemical evolution of gas. This also reduces the computational cost and allows the simulations to be followed for a long period of time.   

\begin{table}[]
\begin{tabular}{|c ||c |c|c |}
\hline
& $\gamma_1$ & $\gamma_2$ & $\gamma_3$ \\ \cline{2-4} 
$\beta$ &  $10^4$-$10^9 \textrm{cm}^{-3}$ & $10^9$-$10^{12}\textrm{cm}^{-3}$ & $10^{12}$-$10^{14} \textrm{cm}^{-3} $   \\ \hline \hline 
(0.0,   0.04) & 1.1362  & 0.9874  & 1.0363  \\ \hline
(0.05, 0.09) & 1.0684  & 1.0865  & 1.0422  \\ \hline
(0.10, 0.12) & 1.0408  & 1.1174  & 1.0591  \\ \hline
(0.13, 0.15) & 1.0214  & 1.1326  & 1.0781  \\ \hline
\end{tabular}
\caption{The values of the polytropic index $\gamma$ used at three density intervals (represented by the number density in cm$^{-3}$) of the collapsed gas for different degrees of rotational support parametrized by $\beta$}
\end{table}

\subsection{Simulation details}
\label{subsec:sim}
Once the central density reaches the critical value given by equation \ref{eqn_ncrit}, the total mass enclosed in single kernel volume $\left(m_{\rm gas}N_{\rm ngb}\right)$ becomes greater than the local Jeans mass which limits the density resolution for SPH simulations. Furthermore, the adaptive time steps for the integration near the critical value become of the order of 0.01 year, which is too small to be able to follow the simulations for any reasonable amount of time after the formation of the central core and hence no fragmentation can be seen. 

To circumvent this problem we search among all the processors for the highest density particle  every ten time steps after the number density($n$) reaches $5 \times 10^{13}  \text{ cm}^{-3}$ for the first time. Since the particles are distributed over a number of processors according to the domain decomposition, we broadcast the information of this highest density particle to all the processors to check for neighbours in their own domain. We then dynamically replace the entire region centralised at the highest density particle with $n \ge 5 \times 10^{14} \text{ cm}^{-3}$ and $T \ge 1300$ K by non gaseous sink particles upon satisfaction of {\it sink formation criteria} as given in \citet{bate97}, i.e., the particle is on current time step and the divergence of both the velocity and acceleration are negative in the vicinity of this particle. The total potential energy within two smoothing lengths is greater than the sum of thermal and rotational kinetic energies and that this region is also virialy unstable. 

The sink particles are formed from about 50 neighbouring gas particles within one smoothing length and thereafter interact with the rest of the gas only gravitationally. A sink can accrete the gas particles falling into an accretion radius $r_{\rm acc}$ that we fix to be about 8 AU. This is done provided the particle is on the current time step, and the total energy of a gas particle relative to the sink is negative i.e. the particle is gravitationally bound to the candidate sink. In addition, the specific angular momentum of the gas particle around the sink is less than what is required to form a circular orbit. 

When a gas particle is accreted, its mass and linear momentum are added to the sink particle and the location of the sink particle is shifted to occupy the centre of mass of the two. The accreted gas particles are removed from the simulation and their effect is taken into account using appropriate boundary conditions near the accretion region. In addition to the accretion radius, we also define an outer accretion radius $r_{\rm outeracc} = 1.25 r_{\rm acc}$ such that the gas particles falling into this outer accretion radius are evolved only gravitationally till they reach the accretion radius and are possibly accreted by the candidate sink. The gas particles may also leave the outer accretion region in the course of their motion. We prevent sink particles to be formed within $2r_{\rm outeracc}$ of each other in order to restrain the counterfeit formation of sink particles from the gas, which eventually would have been accreted by the candidate sink. 

As a check, we also keep track of the global quantities of the gaseous system such as total energy, angular momentum and entropy throughout the simulation. The sink particles are usually created at protostellar density and temperature, and subsequently identified with growing protostars. The gravitational softening for the sink particles is set to be equal to $r_{\rm acc}$ while for the gas particles we use variable gravitational softening length, which is proportional to their SPH smoothing length. This greatly improves the time taken for the simulations to run. Besides, following the discussion in \citet{clark11}, we also implement a constant external pressure boundary in addition to vacuum and periodic boundary conditions in {\sc Gadget}-2.. To this end, we modify the original SPH momentum equation,
\begin{equation}
\frac{dv_i}{dt} = - \sum_{j} m_j \left[ f_i \frac{P_i}{\rho_i^2} \nabla_iW_{ij}(h_i) +  f_j \frac{P_j}{\rho_j^2} \nabla_iW_{ij}(h_j) \right]
\end{equation}
by subtracting the contribution of the external pressure, $P_{\rm ext} = 2.5 \times 10^6 K_B  \text{K} \text{cm}^{-3} $ from both $P_{\rm i}$ and $P_{\rm j}$. All the other symbols have usual meaning. 

\begin{figure}
\centerline{
\includegraphics[width=3.6in]{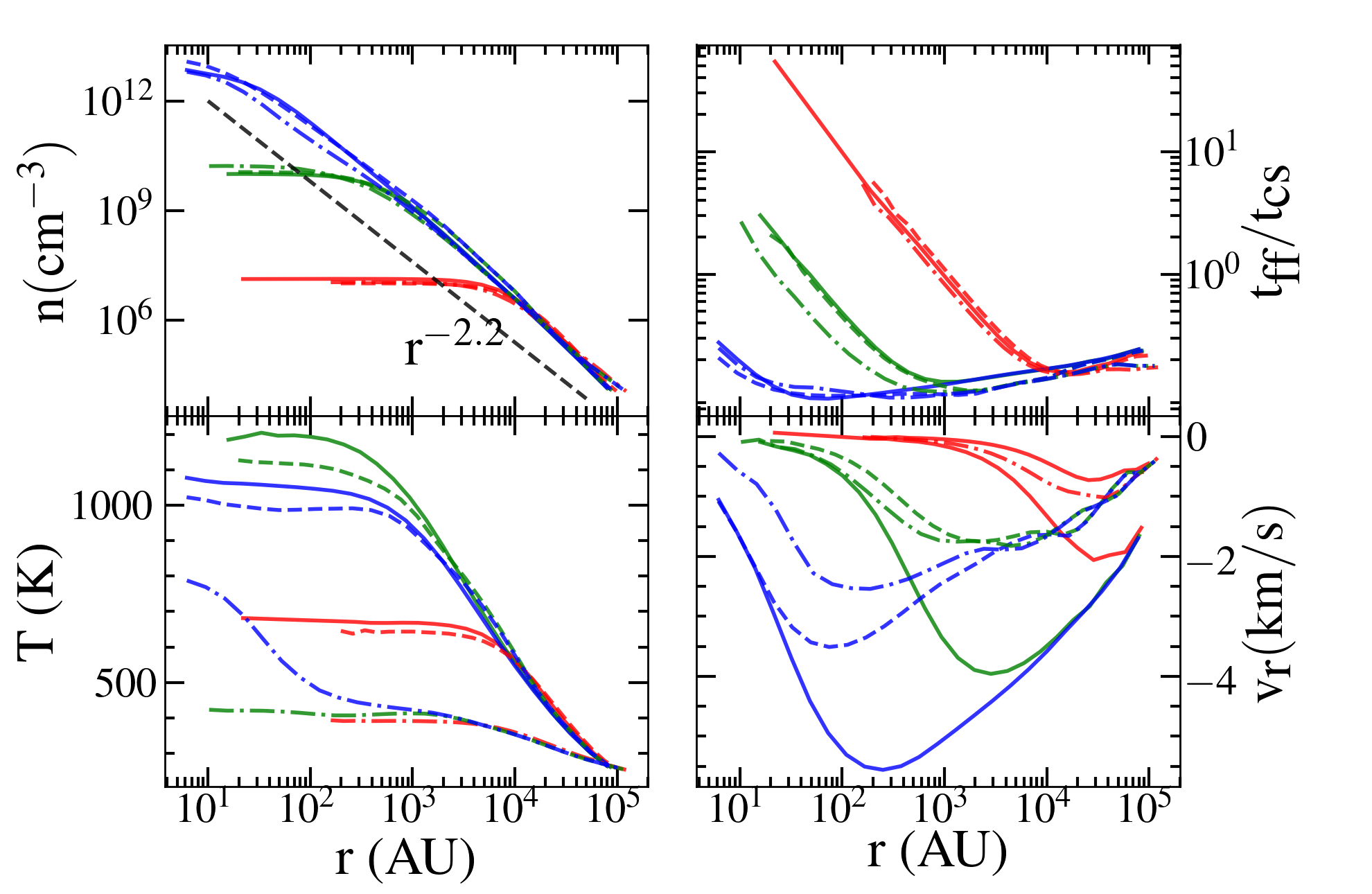}
}
\caption{ Radially binned mass-averaged profile of the number density (top left), free-fall time over sound-crossing time (top right), temperature (bottom left) and the radial velocity (bottom right) in the spherical collapse at various epoch when the central number density reaches a peak $\sim$$10^7$ cm$^{-3}$ (red), $10^{10}$ cm$^{-3}$ (green) and $10^{13}$ cm$^{-3}$ (blue) for the first time. The solid line represents the case with no rotation $\left(\beta_0=0\right)$, dashed for $\beta_0=0.05$ and dot-dashed for $\beta_0=0.1$. The black dashed line in the density plot shows the power-law ($n \propto r^{-2.2}$) dependence. The gas distribution continues to remain self-similar at different epochs during the initial phase of collapse.}
\label{fig:self}
\end{figure}

\subsection{Check for self-similarity solution}
\label{subsec:self}

In this section we quickly check the gas distributions at different epochs of time associated with the runway phase of collapse of the clouds for various degrees of initial solid body rotation. For non rotating $\left(\beta=0\right)$ clouds, the density remains spherically symmetric throughout and follows a well-known power law profile $r^{-2.2}$ while the central density increases monotonically with time as $\propto 1/(t-t_{\rm ff} )^2$. The initial phase of collapse is likely to remain self-similar at different epochs of free-fall time. This means that the collapsing gas distribution is invariant, i.e., looks similar in every scale of density regime. This can be seen from Figure \ref{fig:self}, consistent with the conventional studies \citep{shu77,suto88,on98} in which the gas distribution is a self-similar corresponding to $\gamma_{\rm eff} \sim 1.09$. Clouds with various rotational support also follow roughly the same power-law density profile \citep{matsumoto99,momi08}. However as collapsed gas gets redistributed and accumulated near the centre of mass of the cloud, the degree of rotational support also increases. This causes the density and its gradient to be slightly lower and gas temperatures to be little lower near the centre for clouds with higher degree of rotation \citep{saigo08,by11,meynet13,dutta15}.   

\section{Result}
\label{sec:result}

\begin{figure*}
\centerline{
\includegraphics[width=7.0in]{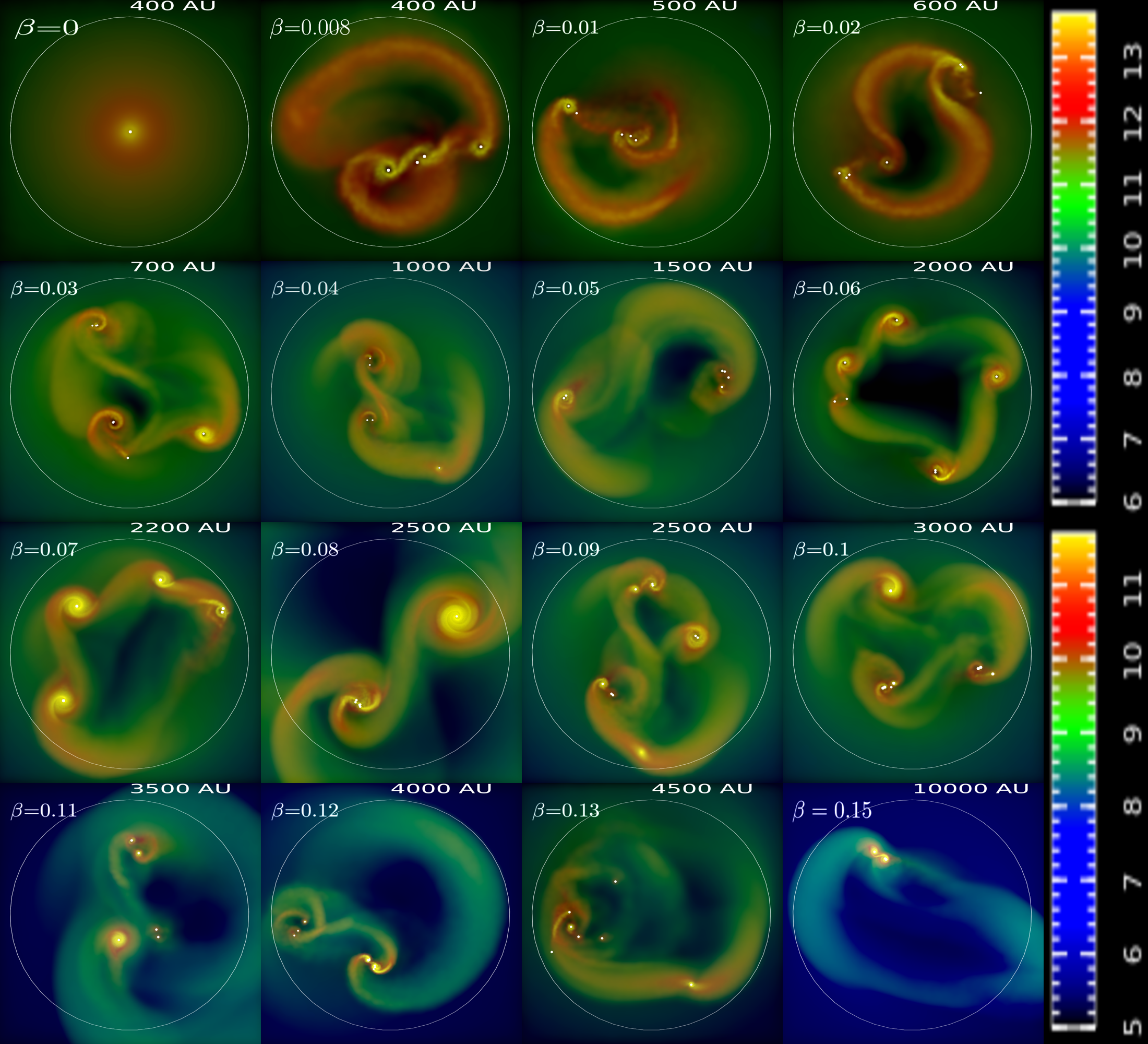}
}
\caption{\label{fig:img} Logarithmically scaled distributions of the number density of the gas clouds in the equatorial plane (plane of rotation of the circumstellar disk) for sixteen realizations $\beta \equiv$ 0.0 - 0.15 at an epoch of time when 50 $M_{\odot}$ has been accreted into the sinks in total. The white dots are the evolving sinks particles in the simulation and the circle shows the size of the central region in AU. As shown in the images, the fast-rotating clouds are likely to develop a small $N$-body system that has a noticeable dense spiral-arms like structure in which protostellar mass evolves substantially on larger length scales as compared to their lower rotating counterparts.}  
\end{figure*}

\begin{figure*}
\centerline{
\includegraphics[width=7.0in]{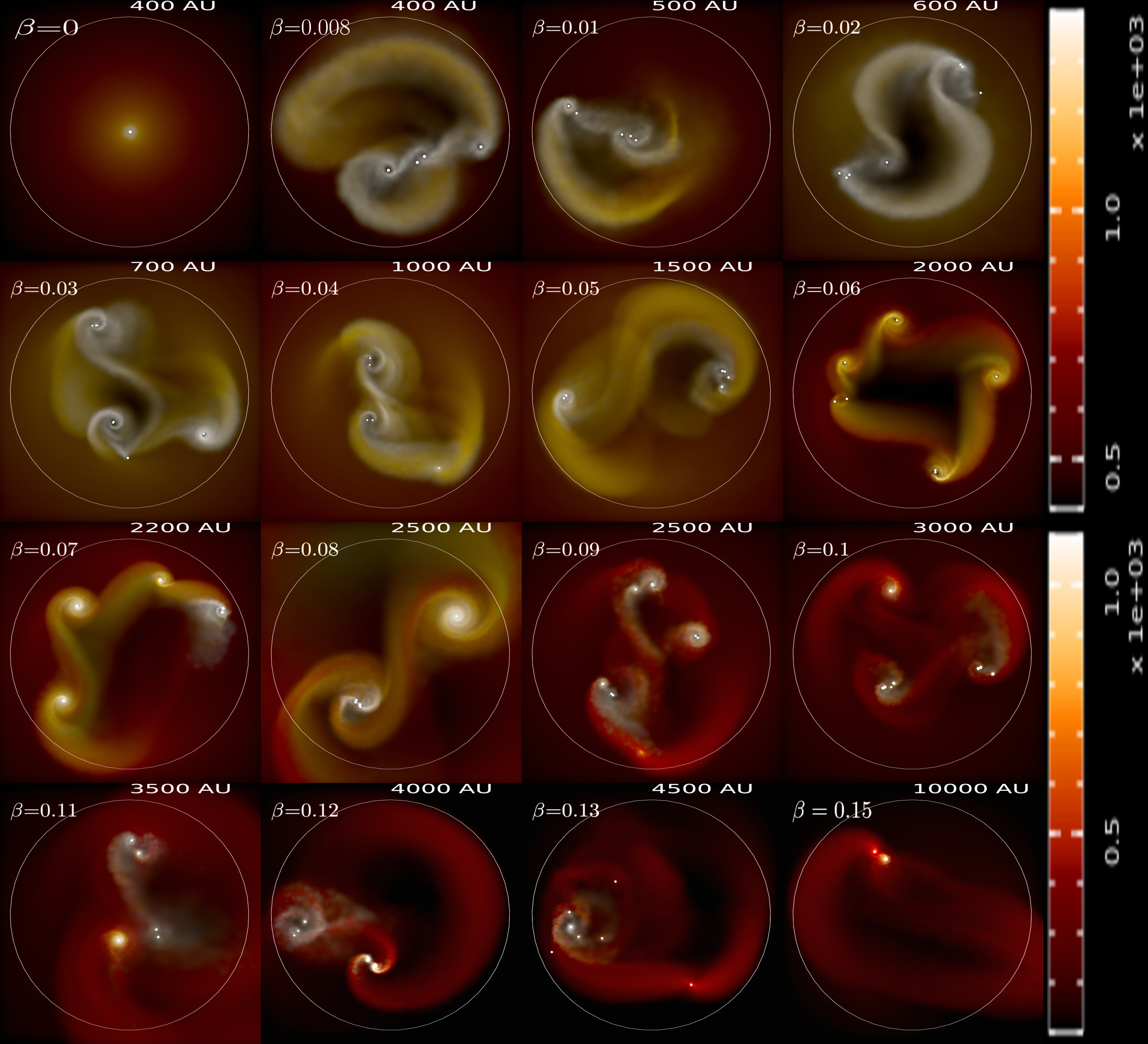}
}
\caption{\label{fig:tmp} Temperature distributions of the gas clouds in the equatorial plane (plane of rotation of the circumstellar disk) for sixteen realizations $\beta \equiv$ 0.0 - 0.15 at an epoch of time when 50 $M_{\odot}$ gas has been accreted into the sinks in total. The evolving sinks are shown by the white dots whereas the circle represents the size of the central region in AU. Since the gas is centrally condensed for slow-rotating clouds,- the temperature near the central region becomes high enough. There is also temperature gradient throughout the layers of the spiral-arms for fast-rotating clouds, which makes them prone to fragmentation due to interplay between gravitational torque and pressure gradient.}  
\end{figure*}
 
The transport of angular momentum to smaller scales results in the formation of rotationally supported spiral-arm, {\it the so-called circumstellar disk or disk-like structures}, around the central hydrostatic core. Till this point the collapse has been studied rigorously \citep{greif12,hirano14,dutta16am,riaz18}. Please see Appendix for runaway collapse phase. Here we follow the simulation further till the epoch of time when $50 M_{\odot}$ of mass are accreted in total onto the dynamically created sink particles as a consequence of instabilities within the spiral-arm and its fragmentation. We study the dynamics associated with multiple sinks and their interaction with ambient gas.

\subsection{Gas distribution during disk fragmentation} 
\label{subsec:fragment}
Figure~\ref{fig:img} \& \ref{fig:tmp} shows the snapshots of logarithmically scaled densities and the temperature distribution of the gas in the equatorial plane of the circumstellar disk for the sixteen values of rotation parameter $\beta$ = 0.0 - 0.15. All the images understandably reflect the fact that the collapsed gas in the spiral-arm becomes unstable by accreting mass from the surrounding and hence prone to fragmentation. This may also be the consequences of interplay between the gravitational torque and pressure gradient throughout the {\it layers} of the spiral-arms. In all these snapshots, the white dots represent the sink particles in the simulation and the circle shows the scale in AU at which the fragmentation takes place. The size of the circle generally keeps on extending with increasing $\beta$-parameter. For example, the central region for $\beta = 0.02$ is approximately 600 AU, whereas for $\beta = 0.1$ it is $\sim$ 3000 AU. The snapshots also reflect the fact that all the clouds with non zero rotation form a small $N$-body protostellar system immersed in the ambient medium near the centre of the clouds. As expected, the fast-rotating clouds are likely to develop noticeable dense spiral-arms in which protostellar mass evolves substantially, and in all probability to fragment reasonably more, say, $N \sim 8 - 11$ for $\beta > 0.05$, as seen in the Fig.~\ref{fig:img} \& \ref{fig:tmp}. This satisfies the justification for conservation of angular momentum during the gas evolution \citep{hirano18}. In contrast, the slow-rotating clouds contain a relatively small number of protostars, $N \sim 4-7$, indicating the possibility of having a high accretion rate, as predicted in theoretical calculations \citep{md13,liu21a}. We have limited our calculations up to this epoch because it is extremely difficult to follow simulations for the fast-rotating clouds above $\beta > 0.15$ beyond this stage of evolution.

Another feature to be noted in Figure \ref{fig:img} is that the gas distribution is highly complicated and non-linear in nature that can be thought of a supersonic and compressible flow coupled to the accreting sinks in the gravitationally bound protostellar system. Due to interaction with the ambient medium, the evolving sinks experiences a strong friction, also known as drag forces \citep{dutta20iau}, which can change the movement and orbit of the sinks \cite[similarly what is seen in the X-ray binaries system][]{bdc17,park21}.

\subsection{Evolution of the sinks} 
\label{subsec:evolution}
\begin{figure*}
\centerline{
\includegraphics[width=7.0in]{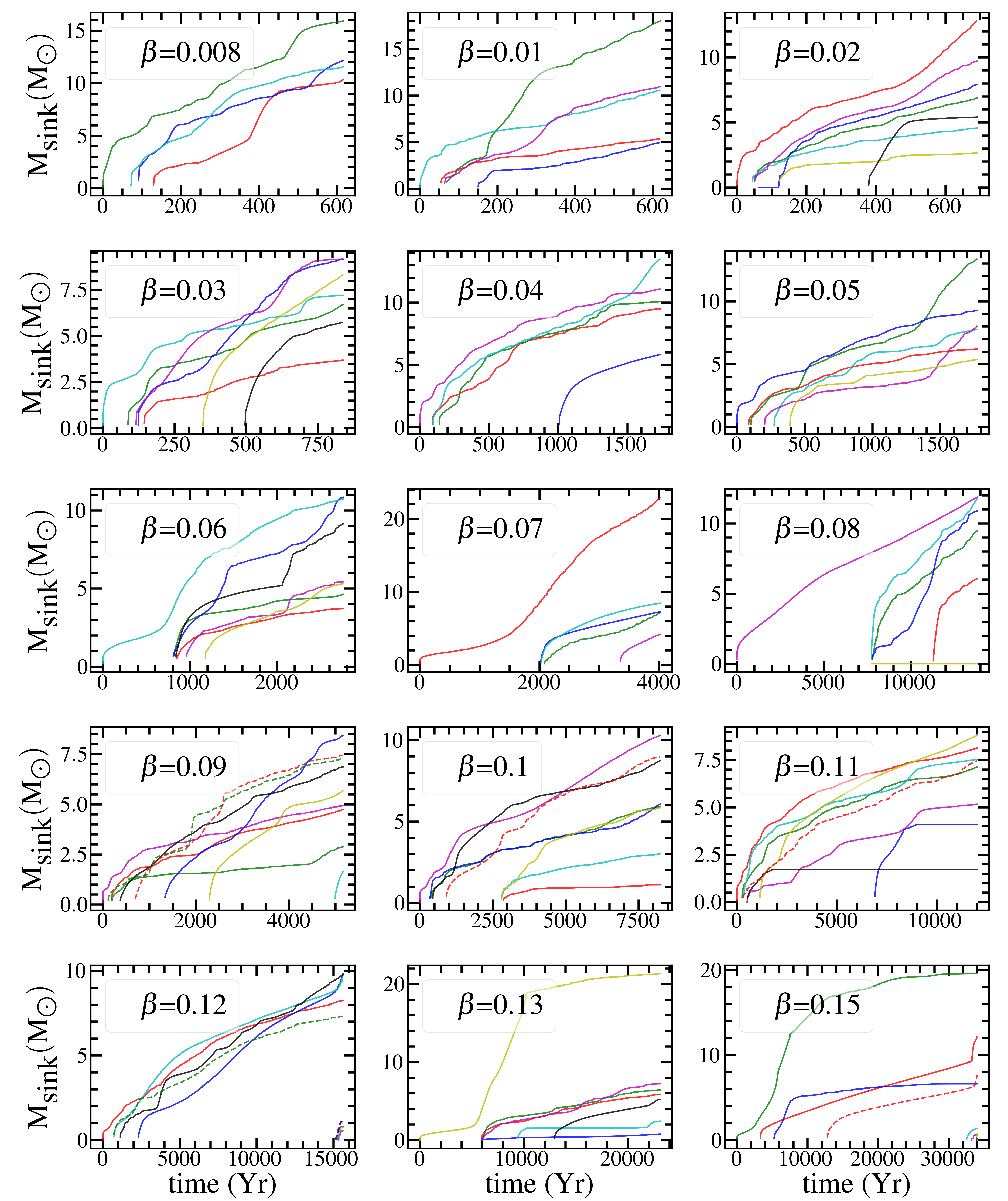}
}
\caption{\label{fig:mass} Mass evolution of the sinks in the simulations are shown as a function of time since the formation of first central hydrostatic core all the way upto the epoch when 50 $M_{\odot}$ has been accreted by the sinks in total, for fifteen instances of the clouds parametrized by $\beta \equiv$ 0.008 - 0.15. Depending on the rotation of the clouds, most of the sinks are formed within a few hundreds to thousands years of evolution from the formation of the central hydrostatic core. Note that fast rotating clouds tend to fragment more to form sinks over a span of mass 0.001 - 20 $M_{\odot}$ depending on the strength of rotational support.}
\end{figure*}

\begin{figure*}
\centerline{
\includegraphics[width=7.0in]{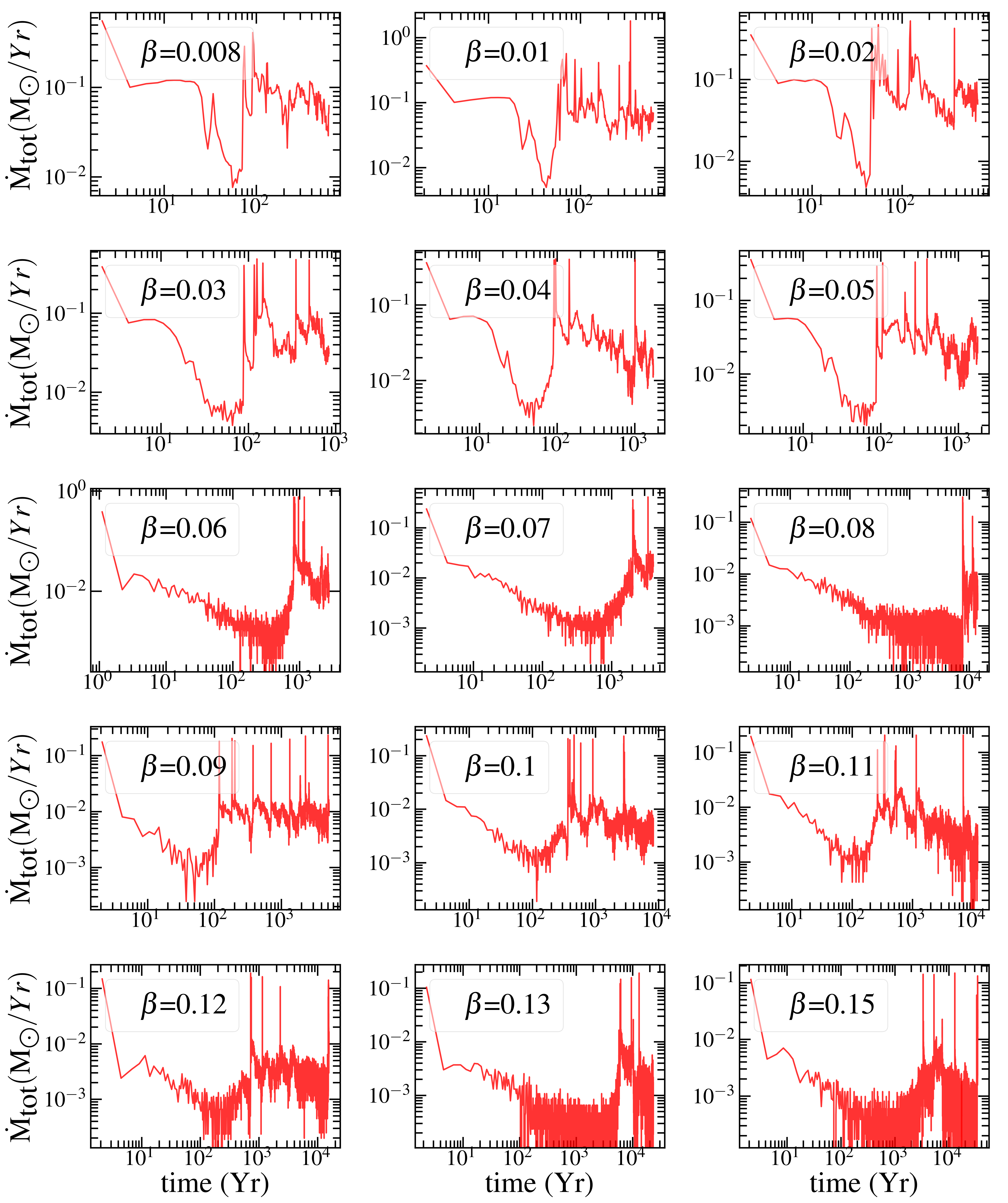}
}
\caption{\label{fig:acc} Different panels show the time evolution of the mass accretion rate by the sinks within the rotating gas clouds quantified by the parameter $\beta$, similar to Figure \ref{fig:mass}, i.e., once roughly $\sum{m_{\rm sink}}\,\sim\,50\,M_{\odot}$ has been accreted by the sinks in total. Sinks formed out of the slow-rotating clouds are expected to accrete the ambient gas at comparatively higher rate (say, $dm/dt \sim 10^{-1} - 10^0$ yr$^{-1}$), whereas accretion for those formed in fast-rotating clouds span over roughly four order of magnitude, and could be as low as $dm/dt \sim 10^{-4}$ yr$^{-1}$.} 
\end{figure*}

The fragmentation of the rotating unstable spiral-arms within the circumstellar disk has significant implications on the final mass of the evolving sinks. Figure \ref{fig:mass} \& \ref{fig:acc} shows the mass of the sinks in our simulation as a function of time. The first feature to be noted is that all clouds are inclined to fragment on an estimated mass scale that evolve up to $\sim$0.001 - 20 $M_{\odot}$ depending on the strength of rotational support of the parent clouds. Because the fragmentation takes place as a consequence of the gravitational instability, the characteristic mass scale may be substantially smaller. Second, as the gas continues to collapse to higher densities, the spiral-arms keep on developing instabilities that heralds in the successive formation of secondary sinks within the circumstellar disk. We see that most of the fragmentation takes place within $\sim$100 - 200 years from the formation of the central core for clouds with $\beta \le 0.05$. Higher is the degree of rotation, longer is the time $t_{\rm frag}$ taken by the gas to become gravitationally unstable to fragmentation. This is expected as the sinks within slow rotating clouds  begin to be Jeans unstable much earlier due to very strong accretion rate, $\sim 1 M_{\odot}/yr$ (as seen in Figure~\ref{fig:acc}). In addition, we see that sinks moving with lower radial velocity within dense ambience are likely to have high accretion rate. Thus, even if the sinks had low mass at the time of formation, their mass can be increased approximately by an order of magnitude relative to their initial mass. As a consequence, sinks now face more gravitational drag due to increase of mass, and therefore they are likely to change their orbits. This also triggers the sinks to be more centrally condensed and continue to accrete gas to end up as massive protostars within a few thousands of years of evolution. We also see that sinks for low  $\beta$-values are quite strongly bound gravitationally. However, clouds with higher degree of rotation tend to fragment more vigorously due to spiral-arm instabilities on larger scales and contain both low and high-mass sinks (some of which even have quite a high radial velocity as compared to the escape velocity of the cloud). For example, a number of protostars with $m_{*} < 1\,M_{\odot}$ are formed for the clouds with $\beta \ge 0.1$. This is consistent with the probability of existence of the smallest fragmentation scale $\sim$$0.03$ AU with $\sim$$0.01\,M_\odot$ \citep{becerra15,hirano17}. We conclude that the formation of the sinks and their dynamical interaction with the ambience depend on the history of collapse (i.e., evolution history of chemical/thermal changes, turbulence and angular momentum conservation). 

\subsection{Histogram of mass function and radial velocity} 
\label{subsec:hist}
\begin{figure*}
\centerline{
\includegraphics[width=7.0in]{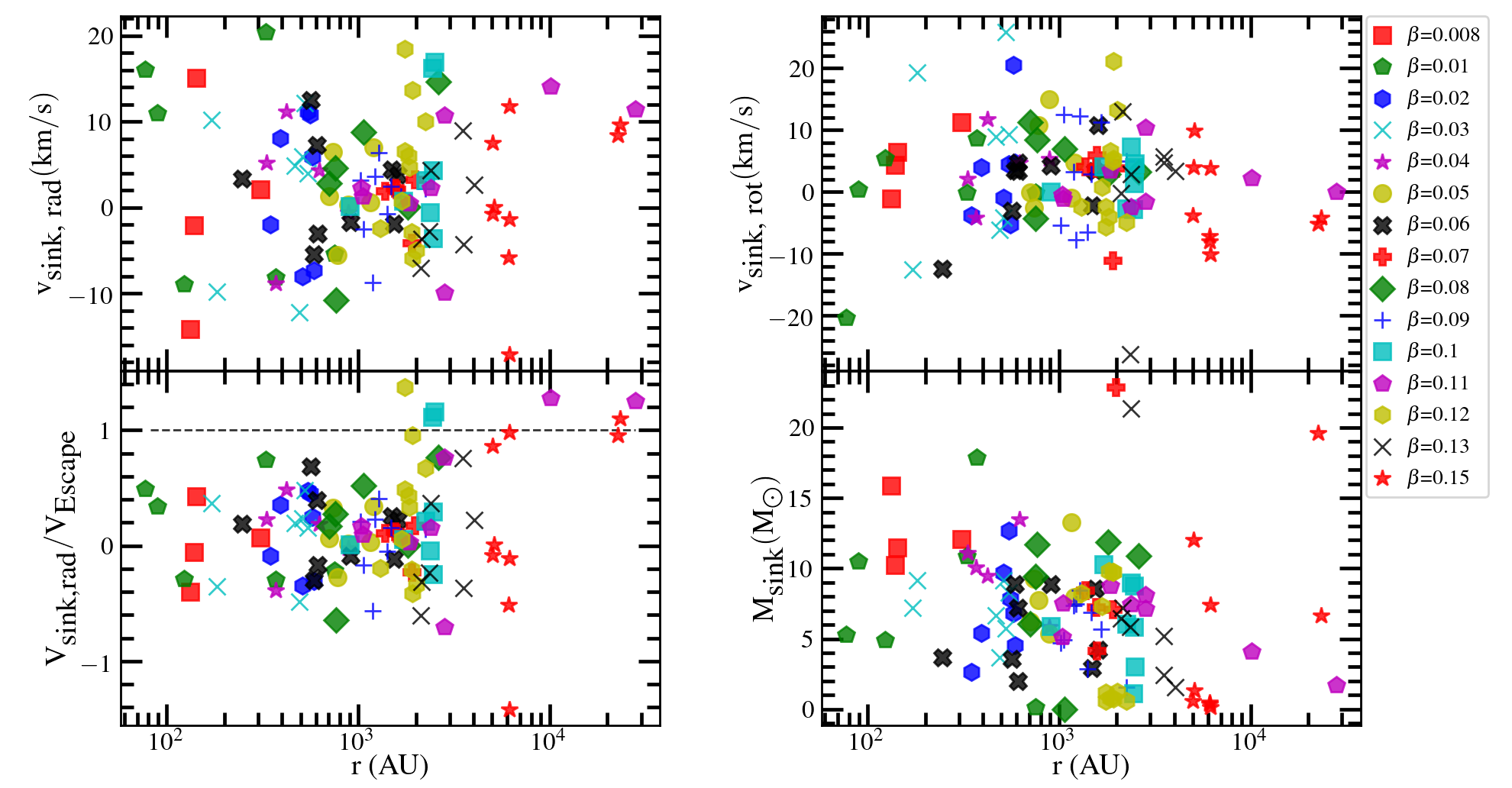}
}
\caption{\label{fig:prop} Properties of the individual evolving sinks formed as a result of disk fragmentation within their parent clouds, parameterised by a wide range of rotational support, are shown in different panel: radial velocity (Top Left), rotational velocity (Top Right), ratio of radial velocity to the escape velocity (Bottom Left) and their mass (Bottom Right) as a function of radial distance from the center of the cloud. This is done at an epoch of time when the multiple system contains a mass roughly $\sum{m_{\rm sink}}\,\sim\,50\,M_{\odot}$. This epoch corresponds to a few thousands of years after the formation of the central hydrostatic core. As discussed in our previous study \citet{dutta16}, it is evident that clouds with higher rotational support tend to fragment more with a range of mass span over low to high, whereas slowly rotating clouds produce comparatively high-mass sinks.}
\end{figure*}

\begin{figure*}
\centerline{
\includegraphics[width=7.0in]{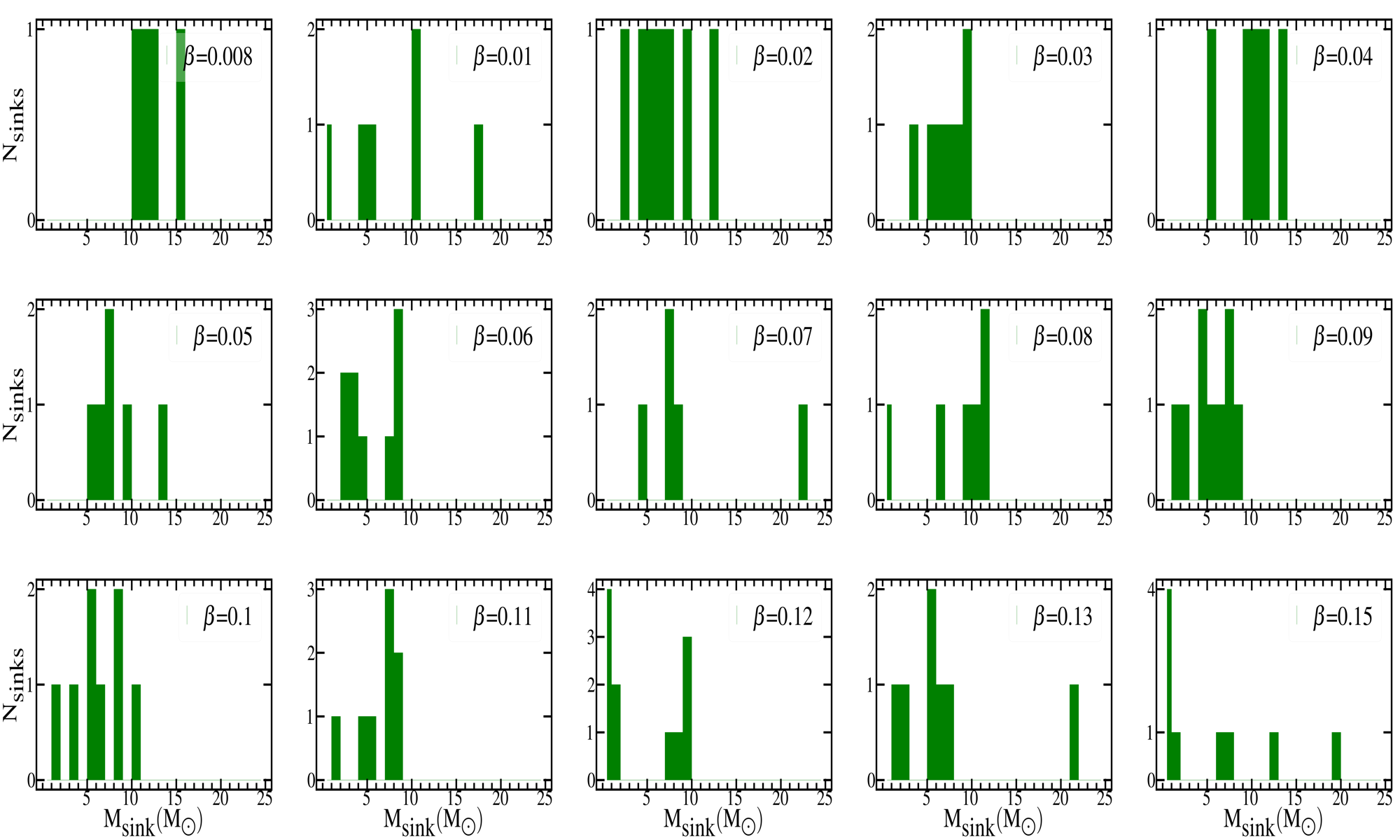}
}
\centerline{
\includegraphics[width=7.0in]{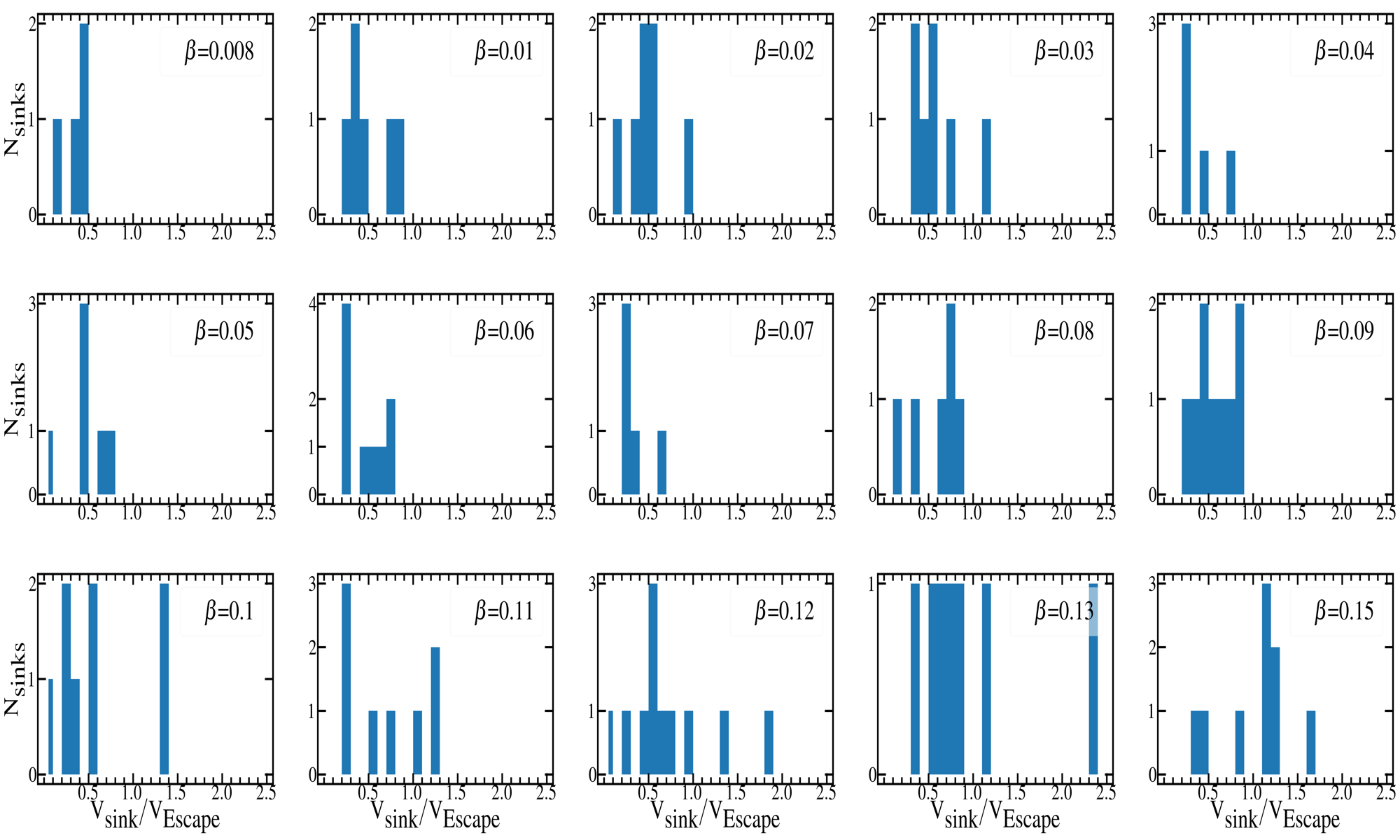}
}
\caption{\label{fig:hist} Various panels show the {\it histograms} for the mass of the sink particles (Top) and the ratio of magnitude of velocities of the sink particle to the escape velocity of the cloud (Bottom) at end of the simulations, i.e., at the epoch when the multi-scaled $N$-body protostellar system attains a total mass of $\sum{m_{\rm sink}}\,\sim\,50\,M_{\odot}$. It is clear that the slow-rotating clouds (e.g., $\beta$ $\equiv$ 0.008, 0.01) are inclined to fragment on a mass scale $\sim$0.001 - 20 $M_{\odot}$ within the central regime $\sim$20 - 100 AU. On the contrary, fast rotating clouds are likely to contain a few low-mass sinks that move faster towards the outer end and accrete a negligible mass from surrounding gaseous medium.} 
\end{figure*}

Here we try to understand the basic properties of the sinks, such as mass, radial velocity and rotational velocity, as reflected in Figure \ref{fig:prop}. On the other hand, Figure \ref{fig:hist} depicts the histogram of the mass function (Top) and the ratio of the radial velocity of the sink particles to the local escape velocity of the cloud (Bottom) at the end of the simulations, i.e., when $50 M_{\odot}$ has been accreted in total. The newly formed sinks in their parent clouds tend to have a wide range of velocities, the typical value of both radial ($v_{\rm rad}$) and rotational ($v_{\rm rot}$) component lies within a span of roughly $\sim$$0.01 - 25$ km s$^{-1}$. This is consistent with the previous studies \cite[e.g.,][]{greif11,dutta16}. In addition, we see that in the absence of radiative feedback, relatively high-mass sinks are likely to be formed irrespective of the rotation of the clouds. Another interesting aspect from the simulations is that a tiny fraction of low-mass sinks for $\beta$ $\equiv$ 0.1 - 0.15 move with relatively high radial velocity as compared to others. They can therefore directly travel at the periphery at later stage of evolution, and can even go away from potential well of the gaseous system with their radial velocity exceeding the escape velocity \cite[$v_{\rm escape} \equiv 10 - 12$~km s$^{-1}$, as seen in][]{dutta20}. Due to high velocity, the ejected sinks can accrete negligible mass from the surrounding. There are subtle deviations between the rotationally supported clouds. From the analysis, it is clear that the mass function for fast rotating clouds peaks at lower mass value, and also the central sink can get quite massive for fast rotating clouds due to larger time scale for fragmentation. The radial velocities for these sinks can also get 2-3 times their escape velocity depending on the strength of the cloud's rotation.

\subsection{Survival possibility}
\label{subsec:survival}
\begin{figure}
	\centerline{
		\includegraphics[width=3.6in]{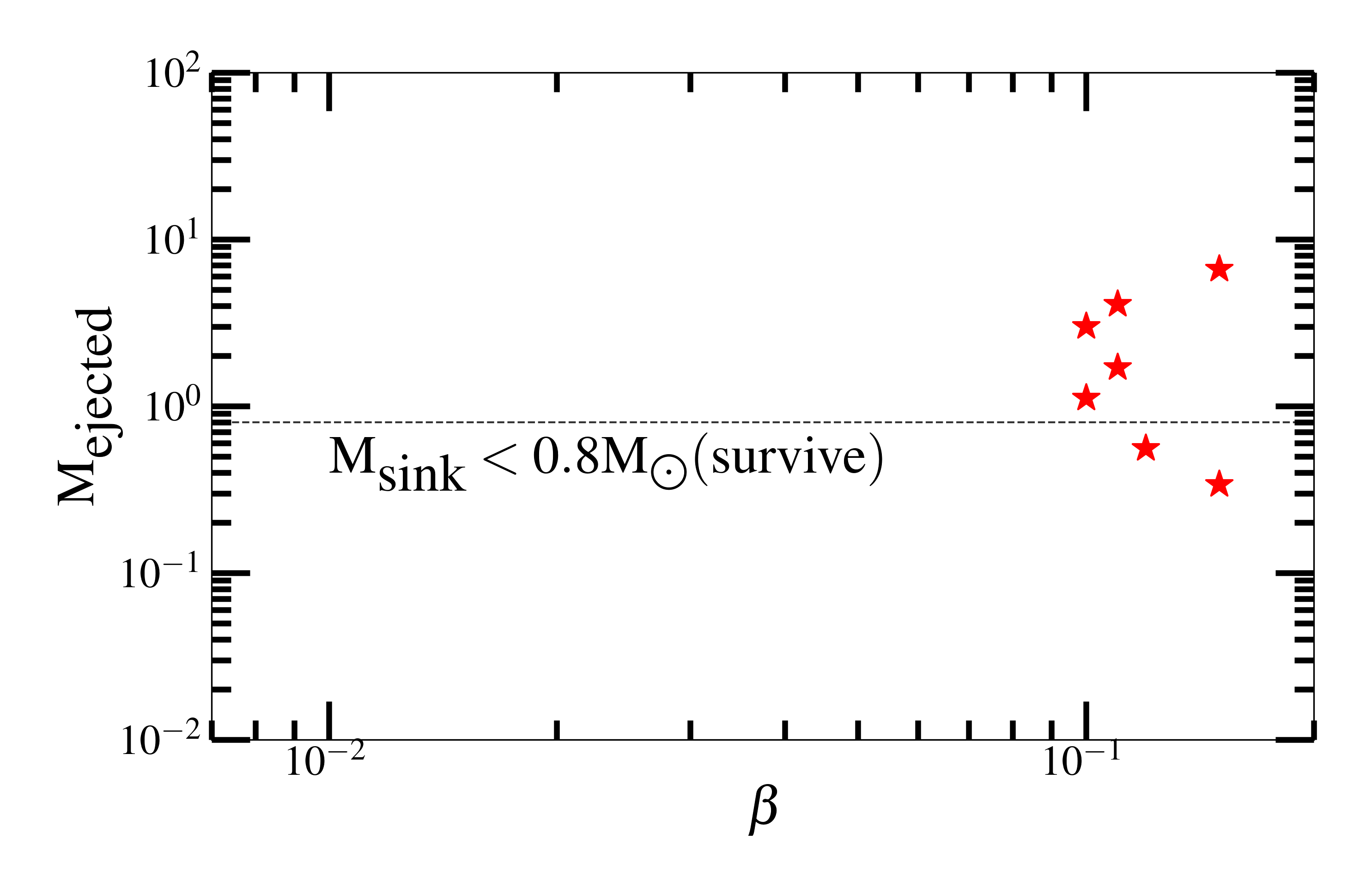}
	}
	\caption{\label{fig:mejected} Mass of the sinks that get ejected from the multi-scaled $N$-body protostellar system is plotted as a function of the rotation parameter $\beta$ of the clouds. The dotted horizontal line indicates the threshold mass($0.8 M_{\odot}$) for the ejected protostars. Only a few sinks are capable of overcoming the strong gravitational drag of the dense gaseous medium and can go away from the potential well of the cluster with radial velocities greater than the escape speed. However, it is noted that only a tiny fraction of them ($\sim$ 10 - 12\%) continue to have a negligible accretion rate for a long period of time to end up as low-mass sinks ($\le 0.8\,M_\odot$). One aspect is that they are likely to remain in the main sequence for billions of years to be survived till present epoch.}      
\end{figure}

Here comes a very important issue to address: the survival possibility of such evolving sinks that may be identified as primordial protostars. Note that the lifetime of a star is inversely proportional to the mass that it contains, say a combination of hydrogen, helium and other higher metallicity gas \cite[see e.g.,][for a detailed calculation of stellar dynamics]{binney08}. In this scenario, a star can survive for billions of years provided its accretion rate remains very minimal so that estimated final mass would be as low as $\sim 0.8$~M$_\odot$ \citep{komiya09,kirihara19}. In Figure \ref{fig:mejected}, we plot
the mass of the sinks that get ejected from the multi-scaled N -body cluster for rotating clouds parametrized
by $\beta$.  As expected, the ejections are likely to happen for the higher rotating gas clouds (e.g. $\beta > 0.1$, as shown in Figure \ref{fig:prop}, the overall mass distribution of the protostars). These realisations show that due to conservation of angular momentum, a fraction of protostars are spread out to the outer periphery of the clouds out of which only a tiny fraction (10 - 12\%) are being ejected as low mass stars. As they continue to have negligible accretion rate for a long period of time, their final mass remains as low as 0.8$M_{\odot}$, and hence likely to survive till today. The dotted horizontal line in Figure \ref{fig:mejected} reflects the threshold mass($0.8 M_{\odot}$) for the ejected protostars. The others can be the massive ones. There is hence a high possibility that a fraction of the sinks within the fast rotating clouds can escape the cluster as low-mass sinks.   


There are considerable chances that they can be evolved as main-sequence stars before entering ZAMS to survive for billions of years till the present epoch. This also confirms the theoretical prediction of the existence probability of first generation stars (it may be either Pop III/Pop II stars or extremely metal poor stars) if they would have contained very low mass ($\le 1$~M$_\odot$) before entering as a ZAMS \citep{andersen09}. Following the recent study by \citet{dutta20} of the Bondi-Hoyle accretion flow, one can estimate the mass-velocity relations for the protostars in which an initial high speed ensures that the mass accretion is relatively smaller. There is hence a good possibility of their existence even in our Galaxy (either in bulge or in halo). Recent state-of-the art observational tools and wide-range surveys have suggested the existence of extremely metal-poor stars that could be the possible candidates of low-mass Pop III/Pop II stars \cite[e.g.][]{komiya10,deb17,husain21}. Therefore searching for more such low-mass stars and where they are located in our Galaxy have become one of the primary interests in present-day observations \citep{johnson15,elbadry18,griffen18,susa19,lb20}.

\section{Summary and Discussion}
\label{sec:summary}

In this work, we performed a suite of 3D simulations using our {\it modified} version of {\sc Gadget}-2 SPH code to follow the gravothermal evolution in a number of primordial gas clouds associated with a degree of rotation that spans over roughly two order of magnitude. The heating and cooling phenomenon that arises during chemical and thermal evolution of the collapsing gas is approximated with a piecewise polytropic equation of state appropriate for the primordial chemistry. Below we outline the main points related to the modification in the SPH simulations and summary of results along with open issues. 

\subsection{Code modification}

\begin{itemize}
\item
In addition to the {\sc ``adiabatic''} and {\sc ``isothermal''} modes available in the publicly available version of {\sc Gadget}-2, we added a {\sc ``polyttropic''} mode in the code, enabling of which in the makefile evolves the gas system with a general polytropic equation of state(EOS). This is done by introducing a polytropic exponent, and appropriately modifying the formulas for internal energy, entropy and rate of change of entropy.

\item
We have carried out sink particle technique in the original {\sc Gadget}-2, following the discussion in \citet{bate97} along with the sink boundary conditions near the accretion surface. We have also defined an outer accretion radius $r_{\rm outeracc} \approx 1.25 r_{\rm acc}$, for the sinks, such that the gas particles falling into this outer accretion radius are evolved only gravitationally. Sink particles can be activated in code by enabling the {\sc sink} mode in the makefile. In this mode, the simulations run till 50 {\it unit of mass} (for example, here we use 50 $M_{\odot}$) has been accreted into the sinks in total, and writes the data related to all the sinks (e.g. mass, position, velocity, internal spin) in their individual text files.
\end{itemize}

\subsection{Summary of results}
\begin{itemize}
\item
Irrespective of the cloud's rotation, most of the sinks start to accrete from the ambient gaseous medium while orbiting the central region. In the absence of radiation mechanism, the continued accretion results in the increase of mass of the evolving sinks, which are likely to turn out to be the high mass protostars, or even massive one, depending on the dynamical evolution. More the mass they accrete, the more the friction they experience due to gravitational drag, which in turn slows their velocity and constrain them to either change their orbital movement or stop. This is more noticeable in slow-rotating clouds where protostars are formed within a few hundreds of AU from the central region and keep on evolving in the absence of radiative feedback from them.

\item
Clouds with higher degree of rotational support tend to fragment more vigorously at far distance from the central region and on larger time scales. After a few thousand years of evolution, a small fraction of these sinks still remain as low mass protostellar objects. In this case, they remain loosely bound to each other gravitationally and may possess radial velocities larger than the escape velocity of the clouds. Thus there is considerable plausibility of these protostars to be ejected from their parent clouds and remain on the main-sequence for a long time. 

\item
The detailed calculation arguably confirms the long-time conjectures of two completely opposite views regarding the mass function of the first generation of stars -- whether they are high-mass or low-mass. Based on our simulations, we have corroborated the fact that the primordial stars might have been formed in a broad range of masses depending on dynamical history of the collapse and initial configuration of the accreting protostars. By looking at the results one can anticipate the fact that the ejected protostars are likely to be survived till present epoch provided their mass remains as low as $m_{\rm Pop III} \le 0.8 M_{\odot}$. The lowest probable mass range of the protostars from our model is comparable to the recent observations \citep{schlaufman18}. 

\end{itemize}

We conclude that the survival rate of a primordial star depends on the initial degree of rotation of the parent gas clouds. Hence simulating long term evolution of the primordial protostellar system is indispensable in order to have an estimation of the initial mass function(IMF) of the Pop III protostars. However, one needs to perform a more rigorous investigation along with inclusion of more sophisticated simulations such as magnetic field, radiative feedback and primordial chemistry.

\subsection{Effect of magnetic field}

A number of studies have confirmed the impact of primordial magnetic fields on the thermodynamics of Pop III star formation \citep{sur10} and on the 21 cm emission line of atomic hydrogen \citep{schleicher09}. Other magneto-hydrodynamics (MHD) simulations \cite[for example,][]{price07,machida08,schleicher2010} show the contribution from so-called magnetic flux \citep{maki07} and magnetic braking \citep{meynet11}. Besides, magnetic fields can also be amplified by orders of magnitudes over their initial cosmological strengths by a combination of small-scale dynamo action and field compression \citep{doi11,sur12,turk12,md13}. Magnetic fields may also provide support against fragmentation \citep{peters14} and outflow \citep{hirano19}. Hence, it is important to note that the inclusion of magnetic fields may substantially influence the gas dynamics, especially the disk evolution and fragmentation phenomenon \cite[see e.g.,][]{mckee20,stacy22}. 

\subsection{Effect of radiative feedback}
The radiation-hydrodynamics (RHD) simulations on the other hand show that the radiation from protostars can significantly change the accretion phenomenon \citep{whalen04,whalen08rad,wise08,wise12,susa09} and can even evaporate the disk \citep{hosokawa11,johnson13,hirano13}. In general we expect feedback to lower the accretion of mass and metals onto protostars \citep{suzuki18} and hence our estimates can be thought of as being upper bounds on these masses \citep{barkana16,barrow17,chon18}. However, it is noted that the radiative feedback becomes important approximately after $\sim$$10^{4}$ year has elapsed from the time of formation of the first protostar. Here, the results from our calculations clearly demonstrate that protostars with $v_{\rm r} \ge v_{\rm esc}$ are able to escape clusters within a few times $10^{4}$ year. Hence they can accrete negligible bulk of mass, which implies that radiation emitted from the surface of these stars are unlikely to have an impact on their mass accretion process. On the other hand, it is obvious that feedback effects need to be included for those protostars roaming around with $v_{\rm r} \ll v_{\rm esc}$. However, analyzing the fate of these protostars is beyond the scope of the present study. 

\subsection{Effect of primordial chemical network}
In a realistic collapse scenario, various chemical species go through the numerous reactions that happen concurrently. This results in the formation of a number of other molecules depending on the local thermodynamic state of the gas \cite[see wonderful reviews by][]{bl_01,cf05}. The thermodynamic balance among these primordial chemical species then determines the net rate of compressional heating and radiative cooling in the gas. Therefore a detailed knowledge of several possible chemical reactions and mass fractions of all the chemical species is required in order to specify the overall chemo-thermal state of the gas \citep{oy03,ripamonti07hd,dutta15}. Hence one has to follow the entire network in detail in order to determine the thermodynamic evolution of the gas, which seems to be fairly complicated to follow in the simulations even for primordial gas with a relatively simpler chemical network of Hydrogen and Helium. 

\subsection{Combined 3D large-scale simulation with Bondi-Hoyle semi-numerical approach}

As discussed above, the protostellar system can numerically be considered as a classical $N$-body problem in which the evolving protostars accretes the ambient gas and coupled with gravity while orbiting around the central gas cloud. However, in reality, this complex phenomenon is indeed difficult to simulate in full 3D simulations \citep{bagla09tree,stacy10} as there is a huge difference of density gradient between the layers of spiral arms. This results in a substantial computational cost. However, with some approximation, the protostellar system can be modeled using semi-analytical calculations with the help of Bondi-Hoyle accretion \citep{bondi44,bondi52} that may provide an satisfactory outcome for the study of stellar dynamics \cite[see e.g.,][] {edgar04,lee14,bdc17,xu19}. It is therefore important to focus on combining the 3D simulation along with Bondi-Hoyle semi-numerical approach \citep{dutta20,park21,keto22} to study the long-term evolution of gas, especially the instabilities that grow during the build-up of the circumstellar disk and complex fragmentation process of its spiral-arms (in preparation). 

\subsection{Evidences for metal-poor stars}

In recent past, a number of studies have predicted that there is a substantial feasibility of finding the first generation of stars, that could be the possible candidate for Pop III stars or extremely metal poor(EMP) stars, in the present-day local universe (\citealt{ws00,tumlinson10,schneider12,gibson13,bsw15,komiya16,ishiyama16}). The search for EMP stars or very metal poor stars have been a cutting-edge research now-a-days, and with the advent of new ultra-modern telescope and futuristic surveys there have been a number of pioneering observational studies that provide a crucial information about the early Universe \citep{dawson04,eisenstein05,tegmark06,fob06,cooke09,bkp09,caffau11,caffau12,sdss12,dawson13,rydberg13,mac13,frebel14,mirocha17}. There have been numerous observational analysis on the evolution of baryonic matter both from cosmological viewpoint \citep{wells21} and statistical interpretation on how the Universe has  evolved into the present state \cite[please see recent studies by][for a preliminary understanding of the evolution]{planck11,planck14,planck16,planck20}. Besides, Pop III stars can also be major origins of both merging binary blackholes and EMP stars, as shown in \citet{tanikawa22}. Recent studies have also shown the possibility of detecting primeval galaxies at higher redshift $z \ge 10$ through JWST observation \citep{jb19,riaz22}. This may constrain the mass function of Pop III stars. Interestingly, Hubble has just detected a very old magnified star of mass $\sim 50 - 100 M_{\odot}$ around redshift 6.2 \citep{welch22}. With the help of state-of-the-art observations from Atacama Large Millimeter/submillimeter Array, \citet{tokuoka22} has also confirmed the possible systematic rotation in the mature stellar population of a $z \sim 9.1$ Galaxy.

\section*{Acknowledgements}

All the simulation are run on the HPSC cluster at Harish-Chandra Research Institute (HRI) at Prayagraj. We thank Jasjeet Singh Bagla, Athena Stacy, Sharanya Sur and Sukalpa Kundu for their constructive comments on the manuscript. This research has made use of NASA's Astrophysics Data System Bibliographic Services.

\appendix{}
\section{Implementation of polytropic equation of state in {\sc Gadget}-2}

The standard model of thermodynamic behaviour of the primordial gas clouds has been well-understood \cite[see for e.g.,][]{abn02,yoha06,gs09,tao09}. The primordial chemical network primarily contains numerous concurrent reactions between hydrogen and helium. For example, at low densities $( n_H \approx 1-10^4 \text{cm}^{-3})$, the hydrogen atoms combine with the free electrons to produce hydrogen molecule ion $H^-$ which in turn combines with the hydrogen atoms to form a small abundance of $H_2$. The gas is then cooled through $H_2$ rotational and vibrational line emission up to temperature of 200K. However, the small abundances of $H_2$ is not sufficient to cool the gas further and the gas begins to heat up with increase in density up to a number density of about $10^8 \text{cm}^{-3}$. At this stage of collapse, the hydrogen atoms are converted to molecules via three body reactions, which again cool the gas through line emissions. The cloud becomes optically thick to the strongest of $H_2$ emission lines beyond number density $10^{11} \text{cm}^{-3}$ \citep{ra04,clark11,dnck15}. However, if we are only interested in the thermodynamic evolution of the gas and not the detailed balance between abundant chemical species, then we can make substantial simplifications and use a general polytropic equation of state for all the chemical processes that involve transfer of heat, \cite[in the line of discussion][]{jappsen05}:

\begin{equation}
T  = a \rho^{\eta-1}\,,
\end{equation}
where $\eta$ is the polytrpoic index. Therefore following the above discussion, we use a polytropic equation of state with a piecewise constant profile for a polytropic index appropriate for thermodynamic evolution of primordial gas clouds. Where the polytropic index $\eta$ changes its values at certain critical number densities, according to the following relations. The publicly available version of {\sc Gadget}-2 can be used to examine the numerically evolved ideal gas system with adiabatic equation of state 
\begin{equation}
P = A \rho^{\gamma} ,\,
\end{equation}      
where $A$ is constant and $\gamma = C_P/C_V $ is the adiabatic index. The internal energy per unit mass, u, of the gas as calculated in the code is
\begin{equation}
u = \left(\frac{A}{\gamma-1 }\right)  \rho^{\gamma -1 } 
\label{inten}
\end{equation}   

In order to model the general polytropic process that can involve the heat transfer as well, we write the following identity from the the first law of thermodynamics 

\begin{equation}
NK_BC \Delta T = P\Delta V + NK_BC_V \Delta T ,\,
\label{firstlaw} 
\end{equation}
where C is a rate of heat added to the system (for adiabatic process C=0), N is the total number of gas particles and the other symbols have their usual meaning. From the above equation and ideal gas law $PV = NK_BT$ we can derive the polytropic equation of state as 
\begin{equation}
P =  B \rho^{\eta} ,\,
\label{poleqs}
\end{equation}      
where $B$ is constant and $\eta = \left( C-C_P \right)/ (C - C_V) $ is the polytropic index. Unlike the adiabatic index $\left(\gamma\right)$ the polytropic index $\left(\eta\right)$ can be greater, smaller or equal to 1, and the two are related by. 

\begin{equation}
\gamma = \eta  + \frac{C}{C_V} \left( 1 - \eta \right)
\end{equation}

Therefore in the code we choose $\gamma = 1 + 1/C_V =  7/5$ for diatomic gas and replace $\gamma$ with $\eta$ at appropriate places, for example the internal energy formula in \ref{inten} is modified to be  
\begin{equation}
u = \left(\frac{B}{\gamma-1 }\right)  \rho^{\eta -1 }
\label{intenmod}
\end{equation}  

With these modification the code can handle general polytropic processes where the temperature can increase $\left( \eta > 1\right)$ or decrease $\left( \eta < 1\right)$ with density and does not require special treatment for the isothermal $\left( \eta = 1\right)$ case.

\section{Runaway collapse phase} 
\begin{figure*}
\centerline{
\includegraphics[width=7.0in]{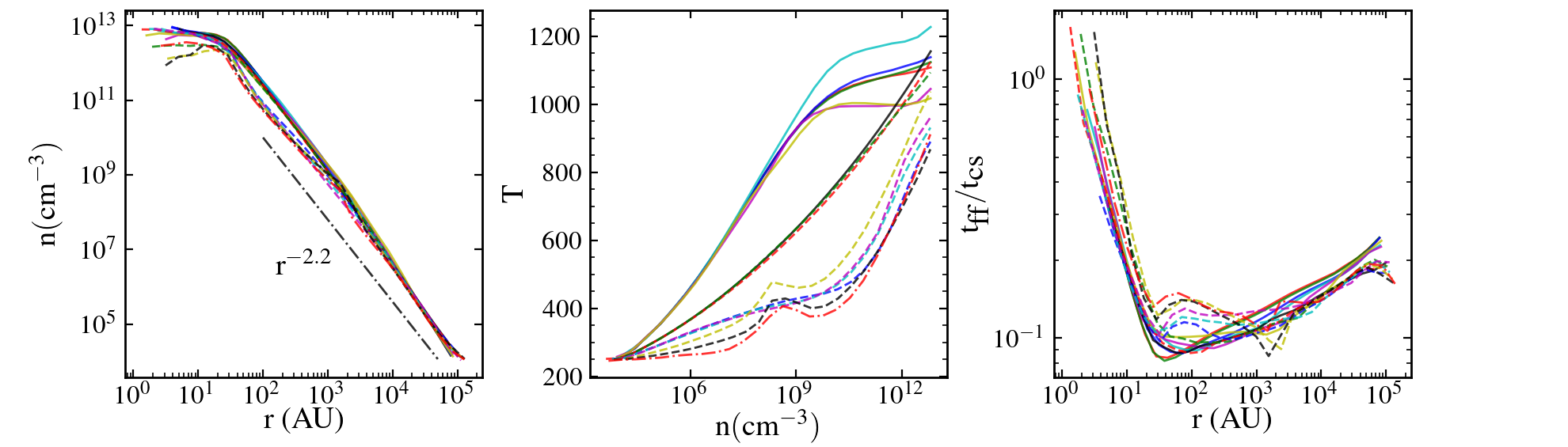}
}
\centerline{
\includegraphics[width=7.0in]{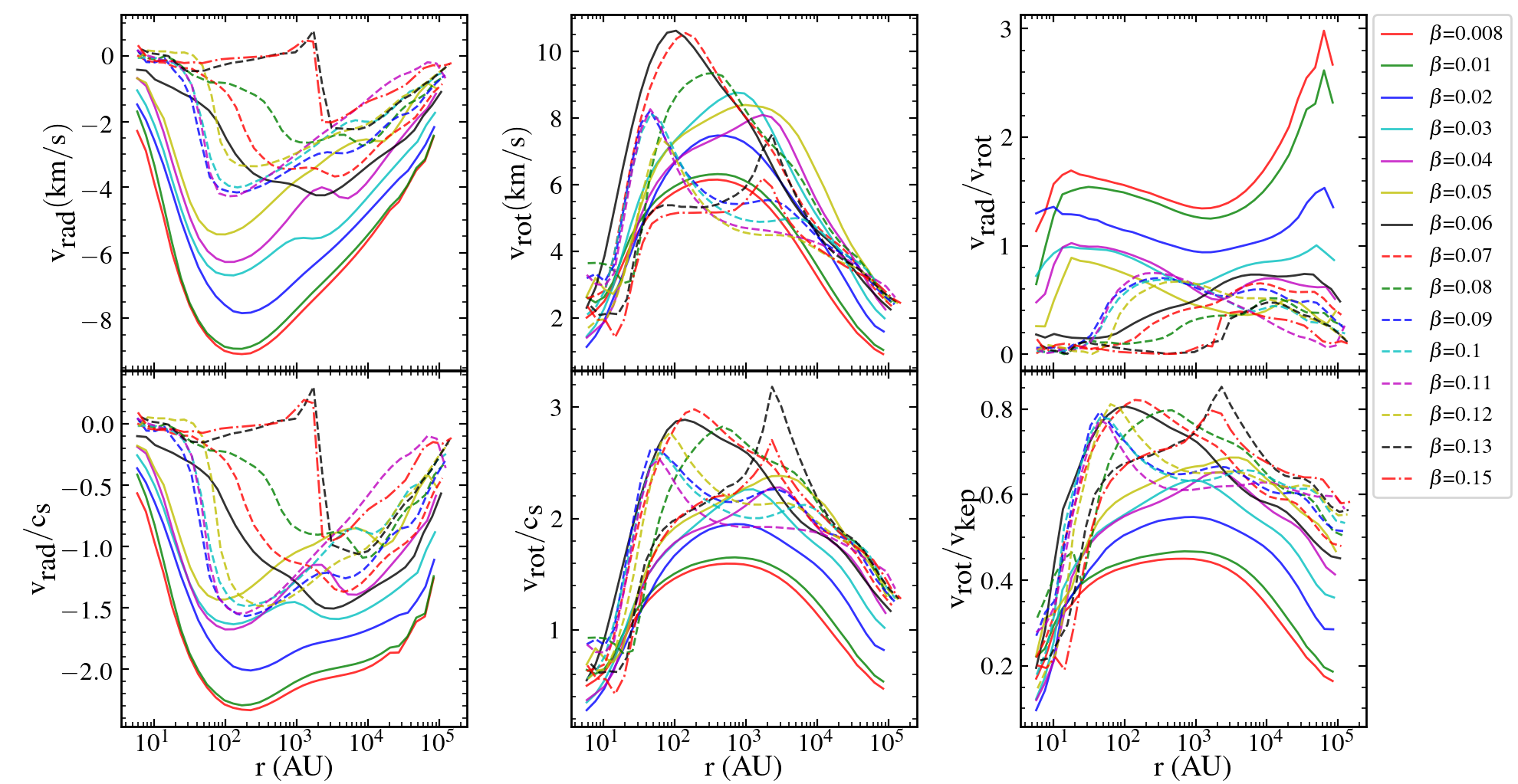}
}
\centerline{
\includegraphics[width=7.0in]{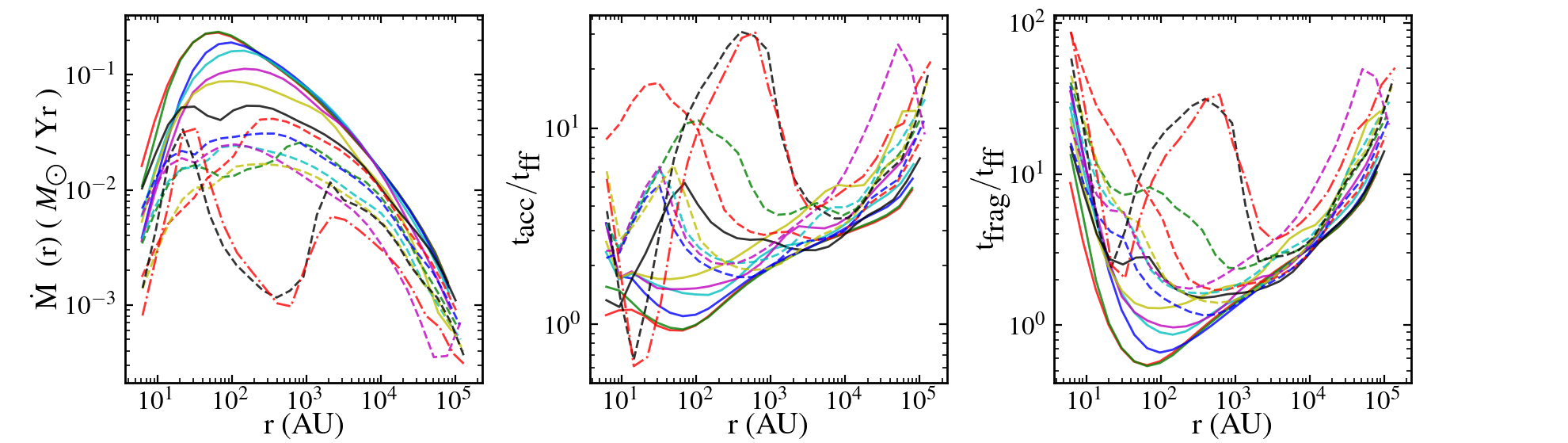}
}
\caption{\label{fig:temp} Initial (runway) collapse phase has been shown for the gas clouds associated with various degrees of initial solid body rotation just before the formation of the central core. The physical properties (estimated as logarithmically scaled radially averaged) of gravitational collapse, such as gas distribution (Top Panel), velocity structure (Middle Panel) and accretion phenomenon (Bottom Panel) are plotted as function of the radius. The collapse remains self similar at different scales and satisfies the {\it power-law} density profile $n \sim r^{-2.2}$ irrespective of the rotation of the clouds, \cite[in agreement of previous studies][]{bl04,g05,yoha06}. It also shows that the slow rotating clouds have higher degree of compressional heating and are therefore hotter compared to fast rotating counterparts. Once the primordial gas gets redistributed near the centre of mass of the cloud, the ratio $v_{\rm rot}(r)/v_{\rm kep}(r)$, where $v_{\rm kep} = \sqrt{GM_{\rm enc}(r)/r}$ and $M_{\rm enc}(r)$ is the mass enclosed inside the sphere of region of radius $r$, indicates that the clouds with higher degree of rotational support go through more efficient phase of angular momentum transport \cite[as shown in][]{greif12,stacy14,dutta16am,hb18}. }
\end{figure*}

In this section we describe the gas distributions, velocity profile and time scales associated with the  initial phase of collapse for the clouds for various degrees of initial solid body rotation till the formation of the central core in our simulations, as illustrated in Figure \ref{fig:temp}. All the physical quantities are radially averaged within logarithmic binned as calculated from our simulations.

The estimated free-fall time over sound crossing time justifies the gravitational collapse of primordial gas, for which the density distribution obey the power-law profile with $n \sim r^{-2.2}$ irrespective of the rotational strength of the clouds. This also confirms that collapse is a self-similar \citep{susa98} process. Therefore the density profile of the collapsed clouds at the outer part is nearly the same as that of the inner regime till the core is formed at the center of clouds \cite[see, e.g.,][for detailed discussion]{meynet13,stacy13,latif15,dutta16am}. Due to high strength of rotation, the radial velocities are considerably lower, as expected for clouds with $\beta \ge 0.05$. So the radial component of velocity is less dominating and gradually becomes comparable to the rotational component near the centre of mass till about 100 AU. This is due to the fact that the infalling gas loses angular momentum near the centre as it gets accreted by the central core. The loss of angular momentum near the centre is compensated by transport of angular momentum farther from the centre. Hence, the angular momentum transport is more noticeable for the clouds with higher degree of rotational support. This is evident from the distribution of the rotational component of the velocities. This also vindicates the formation of larger rotationally supported spiral-arms in the clouds with fast rotating clouds. The flow outside the spiral-arms remains sub keplerian i.e. $v_{\rm rot}(r) \le v_{\rm kep}(r) $, as seen in the middle panel of Figure \ref{fig:temp}. 

In order to quantify the effects of rotation on the accretion phenomenon and associated time scales, we estimate the mass accretion rate, $\dot{M}(r) = 4 \pi r^2 \rho(r) v_{\rm rad}(r)$ and the accretion time $t_{\rm acc} = M_{\rm enc}(r) / 4\pi \rho v_{\rm rad}(r) r^2 $, and plot them as a function of radial distance for different value of the $\beta$-parameter. As can be seen from figure \ref{fig:temp}, the mass accretion rate reaches a maximum value of about 0.1 $M_{\odot}$ per year at about 20 AU from the centre of mass. For distances smaller than this the sound crossing time tends to become comparable to the free fall time which decreases the mass accretion rate near the centre. Because the accretion phenomenon is directly related to the instability within the gas, we also check the degree of Jeans instability in our clouds by measuring the two quantities, $ t_{\rm acc}(r)/t_{\rm ff}(r) $ and $t_{\rm frag}(r)/t_{\rm ff}(r) $ as a function of the radial distance. Here, the fragmentation time, $t_{\rm frag}(r)$, has been estimated by the ratio of Jeans mass and the mass accretion rate of the core \citep{dnck15}. This can provide an approximation of the fragmentation of the gas for different strength of rotation of the clouds. As the infalling gas gets redistributed near the centre of mass of the cloud, the rotational support near the centre also increases, heralding the formation of spiral-arm or disk-like structure. These spiral-arms are likely to become unstable and prone to fragmentation by accreting mass near its boundaries. The fragmentation time scales and sizes of the disks are proportional to their degree of rotation and are considerably larger for fast rotating clouds $\beta \ge 0.05$ than their slowly rotating counterparts for the fixed polytropic index profile.


\begin{thebibliography}{}
\expandafter\ifx\csname natexlab\endcsname\relax\def\natexlab#1{#1}\fi
\providecommand{\url}[1]{\href{#1}{#1}}

\bibitem[{{Abel} {et~al.}(2002){Abel}, {Bryan}, \& {Norman}}]{abn02}
{Abel}, T., {Bryan}, G.~L., \& {Norman}, M.~L. 2002, Science, 295, 93

\bibitem[{{Ahn} {et~al.}(2012){Ahn}, {Alexandroff}, {Allende Prieto},
  {Anderson}, {Anderton}, {Andrews}, {Aubourg}, {Bailey}, {Balbinot}, {Barnes},
  {Bautista}, {Beers}, {Beifiori}, {Berlind}, {Bhardwaj}, {Bizyaev}, {Blake},
  {Blanton}, {Blomqvist}, {Bochanski}, {Bolton}, {Borde}, {Bovy}, {Brandt},
  {Brinkmann}, {Brown}, {Brownstein}, {Bundy}, {Busca}, {Carithers}, {Carnero},
  {Carr}, {Casetti-Dinescu}, {Chen}, {Chiappini}, {Comparat}, {Connolly},
  {Crepp}, {Cristiani}, {Croft}, {Cuesta}, {da Costa}, {Davenport}, {Dawson},
  {de Putter}, {De Lee}, {Delubac}, {Dhital}, {Ealet}, {Ebelke}, {Edmondson},
  {Eisenstein}, {Escoffier}, {Esposito}, {Evans}, {Fan}, {Femen{\'\i}a
  Castell{\'a}}, {Fern{\'a}ndez Alvar}, {Ferreira}, {Filiz Ak}, {Finley},
  {Fleming}, {Font-Ribera}, {Frinchaboy}, {Garc{\'\i}a-Hern{\'a}ndez},
  {Garc{\'\i}a P{\'e}rez}, {Ge}, {G{\'e}nova-Santos}, {Gillespie}, {Girardi},
  {Gonz{\'a}lez Hern{\'a}ndez}, {Grebel}, {Gunn}, {Guo}, {Haggard}, {Hamilton},
  {Harris}, {Hawley}, {Hearty}, {Ho}, {Hogg}, {Holtzman}, {Honscheid},
  {Huehnerhoff}, {Ivans}, {Ivezi{\'c}}, {Jacobson}, {Jiang}, {Johansson},
  {Johnson}, {Kauffmann}, {Kirkby}, {Kirkpatrick}, {Klaene}, {Knapp}, {Kneib},
  {Le Goff}, {Leauthaud}, {Lee}, {Lee}, {Long}, {Loomis}, {Lucatello},
  {Lundgren}, {Lupton}, {Ma}, {Ma}, {MacDonald}, {Mack}, {Mahadevan}, {Maia},
  {Majewski}, {Makler}, {Malanushenko}, {Malanushenko}, {Manchado},
  {Mandelbaum}, {Manera}, {Maraston}, {Margala}, {Martell}, {McBride},
  {McGreer}, {McMahon}, {M{\'e}nard}, {Meszaros}, {Miralda-Escud{\'e}},
  {Montero-Dorta}, {Montesano}, {Morrison}, {Muna}, {Munn}, {Murayama},
  {Myers}, {Neto}, {Nguyen}, {Nichol}, {Nidever}, {Noterdaeme}, {Nuza}, {Ogand
  o}, {Olmstead}, {Oravetz}, {Owen}, {Padmanabhan}, {Palanque-Delabrouille},
  {Pan}, {Parejko}, {Parihar}, {P{\^a}ris}, {Pattarakijwanich}, {Pepper},
  {Percival}, {P{\'e}rez-Fournon}, {P{\'e}rez-R{\`a}fols}, {Petitjean},
  {Pforr}, {Pieri}, {Pinsonneault}, {Porto de Mello}, {Prada}, {Price-Whelan},
  {Raddick}, {Rebolo}, {Rich}, {Richards}, {Robin}, {Rocha-Pinto}, {Rockosi},
  {Roe}, {Ross}, {Ross}, {Rossi}, {Rubi{\~n}o-Martin}, {Samushia}, {Sanchez
  Almeida}, {S{\'a}nchez}, {Santiago}, {Sayres}, {Schlegel}, {Schlesinger},
  {Schmidt}, {Schneider}, {Schultheis}, {Schwope}, {Sc{\'o}ccola}, {Seljak},
  {Sheldon}, {Shen}, {Shu}, {Simmerer}, {Simmons}, {Skibba}, {Skrutskie},
  {Slosar}, {Sobreira}, {Sobeck}, {Stassun}, {Steele}, {Steinmetz}, {Strauss},
  {Streblyanska}, {Suzuki}, {Swanson}, {Tal}, {Thakar}, {Thomas}, {Thompson},
  {Tinker}, {Tojeiro}, {Tremonti}, {Vargas Maga{\~n}a}, {Verde}, {Viel},
  {Vikas}, {Vogt}, {Wake}, {Wang}, {Weaver}, {Weinberg}, {Weiner}, {West},
  {White}, {Wilson}, {Wisniewski}, {Wood-Vasey}, {Yanny}, {Y{\`e}che}, {York},
  {Zamora}, {Zasowski}, {Zehavi}, {Zhao}, {Zheng}, {Zhu}, \& {Zinn}}]{sdss12}
{Ahn}, C.~P., {Alexandroff}, R., {Allende Prieto}, C., {et~al.} 2012, \apjs,
  203, 21

\bibitem[{{Alister Seguel} {et~al.}(2020){Alister Seguel}, {Schleicher},
  {Boekholt}, {Fellhauer}, \& {Klessen}}]{alister20}
{Alister Seguel}, P.~J., {Schleicher}, D.~R.~G., {Boekholt}, T.~C.~N.,
  {Fellhauer}, M., \& {Klessen}, R.~S. 2020, \mnras, 493, 2352

\bibitem[{{Andersen} {et~al.}(2009){Andersen}, {Zinnecker}, {Moneti},
  {McCaughrean}, {Brandl}, {Brandner}, {Meylan}, \& {Hunter}}]{andersen09}
{Andersen}, M., {Zinnecker}, H., {Moneti}, A., {et~al.} 2009, \apj, 707, 1347

\bibitem[{{Bagla} \& {Khandai}(2009)}]{bagla09tree}
{Bagla}, J.~S., \& {Khandai}, N. 2009, \mnras, 396, 2211

\bibitem[{{Bagla} {et~al.}(2009){Bagla}, {Kulkarni}, \& {Padmanabhan}}]{bkp09}
{Bagla}, J.~S., {Kulkarni}, G., \& {Padmanabhan}, T. 2009, \mnras, 397, 971

\bibitem[{{Barkana}(2016)}]{barkana16}
{Barkana}, R. 2016, \physrep, 645, 1

\bibitem[{{Barkana}(2018)}]{barkana18}
---. 2018, \nat, 555, 71

\bibitem[{{Barrow} {et~al.}(2017){Barrow}, {Wise}, {Norman}, {O'Shea}, \&
  {Xu}}]{barrow17}
{Barrow}, K. S.~S., {Wise}, J.~H., {Norman}, M.~L., {O'Shea}, B.~W., \& {Xu},
  H. 2017, \mnras, 469, 4863

\bibitem[{{Bate} \& {Bonnell}(1997)}]{bate97}
{Bate}, M.~R., \& {Bonnell}, I.~A. 1997, \mnras, 285, 33

\bibitem[{{Becerra} {et~al.}(2015){Becerra}, {Greif}, {Springel}, \&
  {Hernquist}}]{becerra15}
{Becerra}, F., {Greif}, T.~H., {Springel}, V., \& {Hernquist}, L.~E. 2015,
  \mnras, 446, 2380

\bibitem[{{Binney} \& {Tremaine}(2008)}]{binney08}
{Binney}, J., \& {Tremaine}, S. 2008, {Galactic Dynamics: Second Edition}

\bibitem[{{Bland-Hawthorn} {et~al.}(2015){Bland-Hawthorn}, {Sutherland}, \&
  {Webster}}]{bsw15}
{Bland-Hawthorn}, J., {Sutherland}, R., \& {Webster}, D. 2015, \apj, 807, 154

\bibitem[{{Bobrick} {et~al.}(2017){Bobrick}, {Davies}, \& {Church}}]{bdc17}
{Bobrick}, A., {Davies}, M.~B., \& {Church}, R.~P. 2017, \mnras, 467, 3556

\bibitem[{Bodenheimer} \& {Boss}(1981)]{bodenheimer81}{Bodenheimer}, P., \& {Boss}, A. P. 1981, \mnras, 197, 477 

\bibitem[{{Bohr} {et~al.}(2020){Bohr}, {Zavala}, {Cyr-Racine}, {Vogelsberger},
  {Bringmann}, \& {Pfrommer}}]{bohr20}
{Bohr}, S., {Zavala}, J., {Cyr-Racine}, F.-Y., {et~al.} 2020, \mnras, 498, 3403

\bibitem[{{Bondi}(1952)}]{bondi52}
{Bondi}, H. 1952, \mnras, 112, 195

\bibitem[{{Bondi} \& {Hoyle}(1944)}]{bondi44}
{Bondi}, H., \& {Hoyle}, F. 1944, \mnras, 104, 273

\bibitem[{{Bovino} {et~al.}(2019){Bovino}, {Schleicher}, \& {Grassi}}]{bsg19}
{Bovino}, S., {Schleicher}, D.~R.~G., \& {Grassi}, T. 2019, Boletin de la
  Asociacion Argentina de Astronomia La Plata Argentina, 61, 274

\bibitem[{{Bromm} \& {Larson}(2004)}]{bl04}
{Bromm}, V., \& {Larson}, R.~B. 2004, \araa, 42, 79

\bibitem[{{Bromm} \& {Yoshida}(2011)}]{by11}
{Bromm}, V., \& {Yoshida}, N. 2011, \araa, 49, 373

\bibitem[{{Caffau} {et~al.}(2011){Caffau}, {Bonifacio}, {Fran{\c{c}}ois},
  {Sbordone}, {Monaco}, {Spite}, {Spite}, {Ludwig}, {Cayrel}, {Zaggia},
  {Hammer}, {Randich}, {Molaro}, \& {Hill}}]{caffau11}
{Caffau}, E., {Bonifacio}, P., {Fran{\c{c}}ois}, P., {et~al.} 2011, \nat, 477,
  67

\bibitem[{{Caffau} {et~al.}(2012){Caffau}, {Bonifacio}, {Fran{\c{c}}ois},
  {Spite}, {Spite}, {Zaggia}, {Ludwig}, {Steffen}, {Mashonkina}, {Monaco},
  {Sbordone}, {Molaro}, {Cayrel}, {Plez}, {Hill}, {Hammer}, \&
  {Randich}}]{caffau12}
---. 2012, \aap, 542, A51

\bibitem[{{Chiaki} \& {Yoshida}(2022)}]{chiaki22}
{Chiaki}, G., \& {Yoshida}, N. 2022, \mnras, 510, 5199

\bibitem[{{Chon} {et~al.}(2018){Chon}, {Hosokawa}, \& {Yoshida}}]{chon18}
{Chon}, S., {Hosokawa}, T., \& {Yoshida}, N. 2018, \mnras, 475, 4104

\bibitem[{{Ciardi} \& {Ferrara}(2005)}]{cf05}
{Ciardi}, B., \& {Ferrara}, A. 2005, \ssr, 116, 625

\bibitem[{{Clark} {et~al.}(2011){Clark}, {Glover}, {Smith}, {Greif}, {Klessen},
  \& {Bromm}}]{clark11}
{Clark}, P.~C., {Glover}, S. C.~O., {Smith}, R.~J., {et~al.} 2011, Science,
  331, 1040

\bibitem[{{Cooke} {et~al.}(2009){Cooke}, {Cooray}, {Chary}, {Bromm}, {Cen},
  {Ellis}, {Fernand ez}, {Furlanetto}, {Loeb}, {Moore}, {Moustakas}, {Oh},
  {O'Shea}, {Scannapieco}, {Smith}, {Trenti}, {Venkatesan}, {Whalen}, \&
  {Yoshida}}]{cooke09}
{Cooke}, J., {Cooray}, A., {Chary}, R.-R., {et~al.} 2009, in astro2010: The
  Astronomy and Astrophysics Decadal Survey, Vol. 2010, 53

\bibitem[{{Dawson} {et~al.}(2013){Dawson}, {Schlegel}, {Ahn}, {Anderson},
  {Aubourg}, {Bailey}, {Barkhouser}, {Bautista}, {Beifiori}, {Berlind},
  {Bhardwaj}, {Bizyaev}, {Blake}, {Blanton}, {Blomqvist}, {Bolton}, {Borde},
  {Bovy}, {Brandt}, {Brewington}, {Brinkmann}, {Brown}, {Brownstein}, {Bundy},
  {Busca}, {Carithers}, {Carnero}, {Carr}, {Chen}, {Comparat}, {Connolly},
  {Cope}, {Croft}, {Cuesta}, {da Costa}, {Davenport}, {Delubac}, {de Putter},
  {Dhital}, {Ealet}, {Ebelke}, {Eisenstein}, {Escoffier}, {Fan}, {Filiz Ak},
  {Finley}, {Font-Ribera}, {G{\'e}nova-Santos}, {Gunn}, {Guo}, {Haggard},
  {Hall}, {Hamilton}, {Harris}, {Harris}, {Ho}, {Hogg}, {Holder}, {Honscheid},
  {Huehnerhoff}, {Jordan}, {Jordan}, {Kauffmann}, {Kazin}, {Kirkby}, {Klaene},
  {Kneib}, {Le Goff}, {Lee}, {Long}, {Loomis}, {Lundgren}, {Lupton}, {Maia},
  {Makler}, {Malanushenko}, {Malanushenko}, {Mandelbaum}, {Manera}, {Maraston},
  {Margala}, {Masters}, {McBride}, {McDonald}, {McGreer}, {McMahon}, {Mena},
  {Miralda-Escud{\'e}}, {Montero-Dorta}, {Montesano}, {Muna}, {Myers},
  {Naugle}, {Nichol}, {Noterdaeme}, {Nuza}, {Olmstead}, {Oravetz}, {Oravetz},
  {Owen}, {Padmanabhan}, {Palanque-Delabrouille}, {Pan}, {Parejko},
  {P{\^a}ris}, {Percival}, {P{\'e}rez-Fournon}, {P{\'e}rez-R{\`a}fols},
  {Petitjean}, {Pfaffenberger}, {Pforr}, {Pieri}, {Prada}, {Price-Whelan},
  {Raddick}, {Rebolo}, {Rich}, {Richards}, {Rockosi}, {Roe}, {Ross}, {Ross},
  {Rossi}, {Rubi{\~n}o-Martin}, {Samushia}, {S{\'a}nchez}, {Sayres}, {Schmidt},
  {Schneider}, {Sc{\'o}ccola}, {Seo}, {Shelden}, {Sheldon}, {Shen}, {Shu},
  {Slosar}, {Smee}, {Snedden}, {Stauffer}, {Steele}, {Strauss}, {Streblyanska},
  {Suzuki}, {Swanson}, {Tal}, {Tanaka}, {Thomas}, {Tinker}, {Tojeiro},
  {Tremonti}, {Vargas Maga{\~n}a}, {Verde}, {Viel}, {Wake}, {Watson}, {Weaver},
  {Weinberg}, {Weiner}, {West}, {White}, {Wood-Vasey}, {Yeche}, {Zehavi},
  {Zhao}, \& {Zheng}}]{dawson13}
{Dawson}, K.~S., {Schlegel}, D.~J., {Ahn}, C.~P., {et~al.} 2013, \aj, 145, 10

\bibitem[{{Dawson} {et~al.}(2004){Dawson}, {Rhoads}, {Malhotra}, {Stern},
  {Dey}, {Spinrad}, {Jannuzi}, {Wang}, \& {Landes}}]{dawson04}
{Dawson}, S., {Rhoads}, J.~E., {Malhotra}, S., {et~al.} 2004, \apj, 617, 707

\bibitem[{{de Bennassuti} {et~al.}(2017){de Bennassuti}, {Salvadori},
  {Schneider}, {Valiante}, \& {Omukai}}]{deb17}
{de Bennassuti}, M., {Salvadori}, S., {Schneider}, R., {Valiante}, R., \&
  {Omukai}, K. 2017, \mnras, 465, 926

\bibitem[{{Doi} \& {Susa}(2011)}]{doi11}
{Doi}, K., \& {Susa}, H. 2011, \apj, 741, 93

\bibitem[{{Dutta}(2015)}]{dutta15}
{Dutta}, J. 2015, \apj, 811, 98

\bibitem[{{Dutta}(2016{\natexlab{a}})}]{dutta16}
---. 2016{\natexlab{a}}, \aap, 585, A59

\bibitem[{{Dutta}(2016{\natexlab{b}})}]{dutta16am}
---. 2016{\natexlab{b}}, \apss, 361, 35

\bibitem[{{Dutta} {et~al.}(2015){Dutta}, {Nath}, {Clark}, \&
  {Klessen}}]{dnck15}
{Dutta}, J., {Nath}, B.~B., {Clark}, P.~C., \& {Klessen}, R.~S. 2015, \mnras,
  450, 202

\bibitem[{{Dutta} {et~al.}(2020{\natexlab{a}}){Dutta}, {Sur}, {Stacy}, \&
  {Bagla}}]{dutta20}
{Dutta}, J., {Sur}, S., {Stacy}, A., \& {Bagla}, J.~S. 2020{\natexlab{a}},
  \apj, 901, 16

\bibitem[{{Dutta} {et~al.}(2020{\natexlab{b}}){Dutta}, {Sur}, {Stacy}, \&
  {Bagla}}]{dutta20iau}
{Dutta}, J., {Sur}, S., {Stacy}, A., \& {Bagla}, J.~S. 2020{\natexlab{b}}, in
  IAU Symposium, Vol. 342, IAU Symposium, ed. K.~{Asada}, E.~{de Gouveia Dal
  Pino}, M.~{Giroletti}, H.~{Nagai}, \& R.~{Nemmen}, 266--267

\bibitem[{{Edgar}(2004)}]{edgar04}
{Edgar}, R. 2004, \nar, 48, 843

\bibitem[{{Eisenstein} {et~al.}(2005){Eisenstein}, {Zehavi}, {Hogg},
  {Scoccimarro}, {Blanton}, {Nichol}, {Scranton}, {Seo}, {Tegmark}, {Zheng},
  {Anderson}, {Annis}, {Bahcall}, {Brinkmann}, {Burles}, {Castand er},
  {Connolly}, {Csabai}, {Doi}, {Fukugita}, {Frieman}, {Glazebrook}, {Gunn},
  {Hendry}, {Hennessy}, {Ivezi{\'c}}, {Kent}, {Knapp}, {Lin}, {Loh}, {Lupton},
  {Margon}, {McKay}, {Meiksin}, {Munn}, {Pope}, {Richmond}, {Schlegel},
  {Schneider}, {Shimasaku}, {Stoughton}, {Strauss}, {SubbaRao}, {Szalay},
  {Szapudi}, {Tucker}, {Yanny}, \& {York}}]{eisenstein05}
{Eisenstein}, D.~J., {Zehavi}, I., {Hogg}, D.~W., {et~al.} 2005, \apj, 633, 560

\bibitem[{{El-Badry} {et~al.}(2018){El-Badry}, {Bland-Hawthorn}, {Wetzel},
  {Quataert}, {Weisz}, {Boylan-Kolchin}, {Hopkins}, {Faucher-Gigu{\`e}re},
  {Kere{\v{s}}}, \& {Garrison-Kimmel}}]{elbadry18}
{El-Badry}, K., {Bland-Hawthorn}, J., {Wetzel}, A., {et~al.} 2018, \mnras, 480,
  652

\bibitem[{{Finkelstein} {et~al.}(2022){Finkelstein}, {Bagley}, {Arrabal Haro},
  {Dickinson}, {Ferguson}, {Kartaltepe}, {Papovich}, {Burgarella}, {Kocevski},
  {Huertas-Company}, {Iyer}, {Larson}, {P{\'e}rez-Gonz{\'a}lez}, {Rose},
  {Tacchella}, {Wilkins}, {Chworowsky}, {Medrano}, {Morales}, {Somerville},
  {Yung}, {Fontana}, {Giavalisco}, {Grazian}, {Grogin}, {Kewley}, {Koekemoer},
  {Kirkpatrick}, {Kurczynski}, {Lotz}, {Pentericci}, {Pirzkal}, {Ravindranath},
  {Ryan}, {Trump}, {Yang}, {Almaini}, {Amor{\'\i}n}, {Annunziatella},
  {Backhaus}, {Barro}, {Behroozi}, {Bell}, {Bhatawdekar}, {Bisigello}, {Bromm},
  {Buat}, {Buitrago}, {Calabr{\'o}}, {Casey}, {Castellano}, {Ch{\'a}vez Ortiz},
  {Ciesla}, {Cleri}, {Cohen}, {Cole}, {Cooke}, {Cooper}, {Cooray}, {Costantin},
  {Cox}, {Croton}, {Daddi}, {Dav{\'e}}, {de la Vega}, {Dekel}, {Elbaz},
  {Estrada-Carpenter}, {Faber}, {Fern{\'a}ndez}, {Finkelstein}, {Freundlich},
  {Fujimoto}, {Garc{\'\i}a-Argum{\'a}nez}, {Gardner}, {Gawiser},
  {G{\'o}mez-Guijarro}, {Guo}, {Hamilton}, {Hathi}, {Holwerda}, {Hirschmann},
  {Hutchison}, {Jha}, {Jogee}, {Juneau}, {Jung}, {Kassin}, {Le Bail}, {Leung},
  {Lucas}, {Magnelli}, {Mantha}, {Matharu}, {McGrath}, {McIntosh}, {Merlin},
  {Mobasher}, {Newman}, {Nicholls}, {Pandya}, {Rafelski}, {Ronayne}, {Santini},
  {Seill{\'e}}, {Shah}, {Shen}, {Simons}, {Snyder}, {Stanway}, {Straughn},
  {Teplitz}, {Vanderhoof}, {Vega-Ferrero}, {Wang}, {Weiner}, {Willmer},
  {Wuyts}, \& {Zavala}}]{finkelstein22}
{Finkelstein}, S.~L., {Bagley}, M.~B., {Arrabal Haro}, P., {et~al.} 2022, arXiv
  e-prints, arXiv:2207.12474

\bibitem[{{Frebel} {et~al.}(2019){Frebel}, {Ji}, {Ezzeddine}, {Hansen},
  {Chiti}, {Thompson}, \& {Merle}}]{frebel19}
{Frebel}, A., {Ji}, A.~P., {Ezzeddine}, R., {et~al.} 2019, \apj, 871, 146

\bibitem[{{Frebel} {et~al.}(2014){Frebel}, {Simon}, \& {Kirby}}]{frebel14}
{Frebel}, A., {Simon}, J.~D., \& {Kirby}, E.~N. 2014, \apj, 786, 74

\bibitem[{{Fukushima} {et~al.}(2020){Fukushima}, {Hosokawa}, {Chiaki},
  {Omukai}, {Yoshida}, \& {Kuiper}}]{fukushima20}
{Fukushima}, H., {Hosokawa}, T., {Chiaki}, G., {et~al.} 2020, \mnras, 497, 829

\bibitem[{{Furlanetto} {et~al.}(2006){Furlanetto}, {Oh}, \& {Briggs}}]{fob06}
{Furlanetto}, S.~R., {Oh}, S.~P., \& {Briggs}, F.~H. 2006, \physrep, 433, 181

\bibitem[{{Gibson} {et~al.}(2013){Gibson}, {Nieuwenhuizen}, \&
  {Schild}}]{gibson13}
{Gibson}, C.~H., {Nieuwenhuizen}, T.~M., \& {Schild}, R.~E. 2013, Journal of
  Cosmology, 22, 10163

\bibitem[{{Glover}(2005)}]{g05}
{Glover}, S. 2005, \ssr, 117, 445

\bibitem[{{Glover} \& {Savin}(2009)}]{gs09}
{Glover}, S.~C.~O., \& {Savin}, D.~W. 2009, \mnras, 393, 911

\bibitem[{{Greif} {et~al.}(2012){Greif}, {Bromm}, {Clark}, {Glover}, {Smith},
  {Klessen}, {Yoshida}, \& {Springel}}]{greif12}
{Greif}, T.~H., {Bromm}, V., {Clark}, P.~C., {et~al.} 2012, \mnras, 424, 399

\bibitem[{{Greif} {et~al.}(2011){Greif}, {Springel}, {White}, {Glover},
  {Clark}, {Smith}, {Klessen}, \& {Bromm}}]{greif11}
{Greif}, T.~H., {Springel}, V., {White}, S. D.~M., {et~al.} 2011, \apj, 737, 75

\bibitem[{{Griffen} {et~al.}(2018){Griffen}, {Dooley}, {Ji}, {O'Shea},
  {G{\'o}mez}, \& {Frebel}}]{griffen18}
{Griffen}, B.~F., {Dooley}, G.~A., {Ji}, A.~P., {et~al.} 2018, \mnras, 474, 443

\bibitem[{{Haemmerl{\'e}} {et~al.}(2020){Haemmerl{\'e}}, {Mayer}, {Klessen},
  {Hosokawa}, {Madau}, \& {Bromm}}]{haemmerle20}
{Haemmerl{\'e}}, L., {Mayer}, L., {Klessen}, R.~S., {et~al.} 2020, \ssr, 216,
  48

\bibitem[{{Hartwig} {et~al.}(2022){Hartwig}, {Magg}, {Chen}, {Tarumi}, {Bromm},
  {Glover}, {Ji}, {Klessen}, {Latif}, {Volonteri}, \& {Yoshida}}]{hartwig22}
{Hartwig}, T., {Magg}, M., {Chen}, L.-H., {et~al.} 2022, arXiv e-prints,
  arXiv:2206.00223

\bibitem[{{Herrington} {et~al.}(2022){Herrington}, {Whalen}, \&
  {Woods}}]{herrington22}
{Herrington}, N.~P., {Whalen}, D.~J., \& {Woods}, T.~E. 2022, arXiv e-prints,
  arXiv:2208.00008

\bibitem[{{Hirano} \& {Bromm}(2018{\natexlab{a}})}]{hirano18}
{Hirano}, S., \& {Bromm}, V. 2018{\natexlab{a}}, \mnras, 476, 3964

\bibitem[{{Hirano} \& {Bromm}(2018{\natexlab{b}})}]{hb18}
---. 2018{\natexlab{b}}, \mnras, 480, L85

\bibitem[{{Hirano} {et~al.}(2017){Hirano}, {Hosokawa}, {Yoshida}, \&
  {Kuiper}}]{hirano17}
{Hirano}, S., {Hosokawa}, T., {Yoshida}, N., \& {Kuiper}, R. 2017, Science,
  357, 1375

\bibitem[{{Hirano} {et~al.}(2014){Hirano}, {Hosokawa}, {Yoshida}, {Umeda},
  {Omukai}, {Chiaki}, \& {Yorke}}]{hirano14}
{Hirano}, S., {Hosokawa}, T., {Yoshida}, N., {et~al.} 2014, \apj, 781, 60

\bibitem[{{Hirano} \& {Machida}(2019)}]{hirano19}
{Hirano}, S., \& {Machida}, M.~N. 2019, \mnras, 485, 4667

\bibitem[{{Hirano} \& {Yoshida}(2013)}]{hirano13}
{Hirano}, S., \& {Yoshida}, N. 2013, \apj, 763, 52

\bibitem[{{Hosokawa} {et~al.}(2011){Hosokawa}, {Omukai}, {Yoshida}, \&
  {Yorke}}]{hosokawa11}
{Hosokawa}, T., {Omukai}, K., {Yoshida}, N., \& {Yorke}, H.~W. 2011, Science,
  334, 1250

\bibitem[{{Husain} {et~al.}(2021){Husain}, {Liu}, \& {Bromm}}]{husain21}
{Husain}, R., {Liu}, B., \& {Bromm}, V. 2021, \mnras, 508, 2169

\bibitem[{{Inoue} \& {Yoshida}(2020)}]{iy20}
{Inoue}, S., \& {Yoshida}, N. 2020, \mnras, 491, L24

\bibitem[{{Ishiyama} {et~al.}(2016){Ishiyama}, {Sudo}, {Yokoi}, {Hasegawa},
  {Tominaga}, \& {Susa}}]{ishiyama16}
{Ishiyama}, T., {Sudo}, K., {Yokoi}, S., {et~al.} 2016, \apj, 826, 9

\bibitem[{{Jappsen} {et~al.}(2005){Jappsen}, {Klessen}, {Larson}, {Li}, \& {Mac
  Low}}]{jappsen05}
{Jappsen}, A.~K., {Klessen}, R.~S., {Larson}, R.~B., {Li}, Y., \& {Mac Low},
  M.~M. 2005, \aap, 435, 611

\bibitem[{{Jeon} \& {Bromm}(2019)}]{jb19}
{Jeon}, M., \& {Bromm}, V. 2019, \mnras, 485, 5939

\bibitem[{{Jeon} {et~al.}(2021){Jeon}, {Bromm}, {Besla}, {Yoon}, \&
  {Choi}}]{jeon21}
{Jeon}, M., {Bromm}, V., {Besla}, G., {Yoon}, J., \& {Choi}, Y. 2021, \mnras,
  502, 1

\bibitem[{{Johnson}(2015)}]{johnson15}
{Johnson}, J.~L. 2015, \mnras, 453, 2771

\bibitem[{{Johnson} {et~al.}(2013){Johnson}, {Dalla Vecchia}, \&
  {Khochfar}}]{johnson13}
{Johnson}, J.~L., {Dalla Vecchia}, C., \& {Khochfar}, S. 2013, \mnras, 428,
  1857

\bibitem[{{Keto} \& {Kuiper}(2022)}]{keto22}
{Keto}, E., \& {Kuiper}, R. 2022, \mnras, 509, 559

\bibitem[{{Kirihara} {et~al.}(2019){Kirihara}, {Tanikawa}, \&
  {Ishiyama}}]{kirihara19}
{Kirihara}, T., {Tanikawa}, A., \& {Ishiyama}, T. 2019, \mnras, 486, 5917

\bibitem[{{Klessen}(2019)}]{klessen19}
{Klessen}, R. 2019, {Formation of the first stars}, ed. M.~{Latif} \&
  D.~{Schleicher}, 67--97

\bibitem[{{Komiya} {et~al.}(2009){Komiya}, {Habe}, {Suda}, \&
  {Fujimoto}}]{komiya09}
{Komiya}, Y., {Habe}, A., {Suda}, T., \& {Fujimoto}, M.~Y. 2009, \apjl, 696,
  L79

\bibitem[{{Komiya} {et~al.}(2010){Komiya}, {Habe}, {Suda}, \&
  {Fujimoto}}]{komiya10}
---. 2010, \apj, 717, 542

\bibitem[{{Komiya} {et~al.}(2016){Komiya}, {Suda}, \& {Fujimoto}}]{komiya16}
{Komiya}, Y., {Suda}, T., \& {Fujimoto}, M.~Y. 2016, \apj, 820, 59

\bibitem[{{Krumholz} \& {McKee}(2005)}]{Krumholz05}
{Krumholz} M. R., {McKee} C. F., 2005, ApJ, 630, 250

\bibitem[{{Kulkarni} {et~al.}(2019){Kulkarni}, {Visbal}, \&
  {Bryan}}]{kulkarni19}
{Kulkarni}, M., {Visbal}, E., \& {Bryan}, G.~L. 2019, \apj, 882, 178


\bibitem[{{Larson}(1972)}]{larson72}
{Larson}, R.~B. 1972, \mnras, 156, 437

\bibitem[{{Larson}(1984)}]{larson84}
{Larson}, R.~B. 1984, \mnras, 206, 197

\bibitem[{{Latif} \& {Schleicher}(2015)}]{latif15}
{Latif}, M.~A., \& {Schleicher}, D.~R.~G. 2015, \mnras, 449, 77

\bibitem[{{Latif} {et~al.}(2022){Latif}, {Whalen}, \& {Khochfar}}]{latif22}
{Latif}, M.~A., {Whalen}, D., \& {Khochfar}, S. 2022, \apj, 925, 28

\bibitem[{{Lee} {et~al.}(2014){Lee}, {Cunningham}, {McKee}, \& {Klein}}]{lee14}
{Lee}, A.~T., {Cunningham}, A.~J., {McKee}, C.~F., \& {Klein}, R.~I. 2014,
  \apj, 783, 50

\bibitem[{{Liu} \& {Bromm}(2020)}]{lb20}
{Liu}, B., \& {Bromm}, V. 2020, arXiv e-prints, arXiv:2006.15260

\bibitem[{{Liu} {et~al.}(2021){Liu}, {Meynet}, \& {Bromm}}]{liu21a}
{Liu}, B., {Meynet}, G., \& {Bromm}, V. 2021, \mnras, 501, 643

\bibitem[{{Loeb} \& {Barkana}(2001)}]{bl_01}
{Loeb}, A., \& {Barkana}, R. 2001, \araa, 39, 19

\bibitem[{{MacDonald} {et~al.}(2013){MacDonald}, {Lawlor}, {Anilmis}, \&
  {Rufo}}]{mac13}
{MacDonald}, J., {Lawlor}, T.~M., {Anilmis}, N., \& {Rufo}, N.~F. 2013, \mnras,
  431, 1425

\bibitem[{{Machida} \& {Doi}(2013)}]{md13}
{Machida}, M.~N., \& {Doi}, K. 2013, \mnras, 435, 3283

\bibitem[{{Machida} {et~al.}(2008{\natexlab{a}}){Machida}, {Matsumoto}, \&
  {Inutsuka}}]{machida08}
{Machida}, M.~N., {Matsumoto}, T., \& {Inutsuka}, S.-i. 2008{\natexlab{a}},
  \apj, 685, 690

\bibitem[{{Machida} {et~al.}(2008{\natexlab{b}}){Machida}, {Omukai},
  {Matsumoto}, \& {Inutsuka}}]{momi08}
{Machida}, M.~N., {Omukai}, K., {Matsumoto}, T., \& {Inutsuka}, S.-i.
  2008{\natexlab{b}}, \apj, 677, 813

\bibitem[{{Madau} \& {Rees}(2001)}]{mr01}
{Madau}, P., \& {Rees}, M.~J. 2001, \apjl, 551, L27

\bibitem[{{Maki} \& {Susa}(2007)}]{maki07}
{Maki}, H., \& {Susa}, H. 2007, \pasj, 59, 787

\bibitem[{{Marigo} {et~al.}(2001){Marigo}, {Girardi}, {Chiosi}, \&
  {Wood}}]{marigo01}
{Marigo}, P., {Girardi}, L., {Chiosi}, C., \& {Wood}, P.~R. 2001, \aap, 371,
  152

\bibitem[{{Matsumoto} \& {Hanawa}(1999)}]{matsumoto99}
{Matsumoto}, T., \& {Hanawa}, T. 1999, \apj, 521, 659

\bibitem[{{Matsumoto} {et~al.}(2015){Matsumoto}, {Nakauchi}, {Ioka}, {Heger},
  \& {Nakamura}}]{matsumoto15}
{Matsumoto}, T., {Nakauchi}, D., {Ioka}, K., {Heger}, A., \& {Nakamura}, T.
  2015, \apj, 810, 64

\bibitem[{{May} \& {Springel}(2021)}]{may21}
{May}, S., \& {Springel}, V. 2021, \mnras, 506, 2603

\bibitem[{{McKee} \& {Tan}(2002)}]{McKee02}
{McKee} C. F., \& {Tan} J. C., 2002, \nat, 416, 59

\bibitem[{{McKee} {et~al.}(2020){McKee}, {Stacy}, \& {Li}}]{mckee20}
{McKee}, C.~F., {Stacy}, A., \& {Li}, P.~S. 2020, \mnras, 496, 5528

\bibitem[{{Meynet} {et~al.}(2011){Meynet}, {Eggenberger}, \&
  {Maeder}}]{meynet11}
{Meynet}, G., {Eggenberger}, P., \& {Maeder}, A. 2011, \aap, 525, L11

\bibitem[{{Meynet} {et~al.}(2013){Meynet}, {Ekstrom}, {Maeder}, {Eggenberger},
  {Saio}, {Chomienne}, \& {Haemmerl{\'e}}}]{meynet13}
{Meynet}, G., {Ekstrom}, S., {Maeder}, A., {et~al.} 2013, {Models of Rotating
  Massive Stars: Impacts of Various Prescriptions}, ed. M.~{Goupil},
  K.~{Belkacem}, C.~{Neiner}, F.~{Ligni{\`e}res}, \& J.~J. {Green}, Vol. 865, 3

\bibitem[{{Meynet} \& {Maeder}(2002)}]{meynet02}
{Meynet}, G., \& {Maeder}, A. 2002, \aap, 390, 561

\bibitem[{{Mirocha} {et~al.}(2017){Mirocha}, {Furlanetto}, \&
  {Sun}}]{mirocha17}
{Mirocha}, J., {Furlanetto}, S.~R., \& {Sun}, G. 2017, \mnras, 464, 1365

\bibitem[{{Omukai} \& {Nishi}(1998)}]{on98}
{Omukai}, K., \& {Nishi}, R. 1998, \apj, 508, 141

\bibitem[{{Omukai} \& {Yoshii}(2003)}]{oy03}
{Omukai}, K., \& {Yoshii}, Y. 2003, \apj, 599, 746

\bibitem[{{Pallottini} {et~al.}(2017){Pallottini}, {Ferrara}, {Bovino},
  {Vallini}, {Gallerani}, {Maiolino}, \& {Salvadori}}]{pallottini17}
{Pallottini}, A., {Ferrara}, A., {Bovino}, S., {et~al.} 2017, \mnras, 471, 4128

\bibitem[{{Park} {et~al.}(2021){Park}, {Ricotti}, \& {Sugimura}}]{park21}
{Park}, J., {Ricotti}, M., \& {Sugimura}, K. 2021, \mnras, 508, 6176

\bibitem[{{Peters} {et~al.}(2014){Peters}, {Schleicher}, {Smith}, {Schmidt}, \&
  {Klessen}}]{peters14}
{Peters}, T., {Schleicher}, D. R.~G., {Smith}, R.~J., {Schmidt}, W., \&
  {Klessen}, R.~S. 2014, \mnras, 442, 3112

\bibitem[{{Planck Collaboration} {et~al.}(2011){Planck Collaboration}, {Ade},
  {Aghanim}, {Arnaud}, {Ashdown}, {Aumont}, {Baccigalupi}, {Baker}, {Balbi},
  {Banday}, {Barreiro}, {Bartlett}, {Battaner}, {Benabed}, {Bennett},
  {Beno{\^\i}t}, {Bernard}, {Bersanelli}, {Bhatia}, {Bock}, {Bonaldi}, {Bond},
  {Borrill}, {Bouchet}, {Bradshaw}, {Bremer}, {Bucher}, {Burigana}, {Butler},
  {Cabella}, {Cantalupo}, {Cappellini}, {Cardoso}, {Carr}, {Casale},
  {Catalano}, {Cay{\'o}n}, {Challinor}, {Chamballu}, {Charra}, {Chary},
  {Chiang}, {Chiang}, {Christensen}, {Clements}, {Colombi}, {Couchot},
  {Coulais}, {Crill}, {Crone}, {Crook}, {Cuttaia}, {Danese}, {D'Arcangelo},
  {Davies}, {Davis}, {de Bernardis}, {de Bruin}, {de Gasperis}, {de Rosa}, {de
  Zotti}, {Delabrouille}, {Delouis}, {D{\'e}sert}, {Dick}, {Dickinson},
  {Dolag}, {Dole}, {Donzelli}, {Dor{\'e}}, {D{\"o}rl}, {Douspis}, {Dupac},
  {Efstathiou}, {En{\ss}lin}, {Eriksen}, {Finelli}, {Foley}, {Forni},
  {Fosalba}, {Frailis}, {Franceschi}, {Freschi}, {Gaier}, {Galeotta},
  {Gallegos}, {Gandolfo}, {Ganga}, {Giard}, {Giardino}, {Gienger},
  {Giraud-H{\'e}raud}, {Gonz{\'a}lez}, {Gonz{\'a}lez-Nuevo}, {G{\'o}rski},
  {Gratton}, {Gregorio}, {Gruppuso}, {Guyot}, {Haissinski}, {Hansen},
  {Harrison}, {Helou}, {Henrot-Versill{\'e}}, {Hern{\'a}ndez-Monteagudo},
  {Herranz}, {Hildebrandt}, {Hivon}, {Hobson}, {Holmes}, {Hornstrup}, {Hovest},
  {Hoyland}, {Huffenberger}, {Jaffe}, {Jagemann}, {Jones}, {Juillet}, {Juvela},
  {Kangaslahti}, {Keih{\"a}nen}, {Keskitalo}, {Kisner}, {Kneissl}, {Knox},
  {Krassenburg}, {Kurki-Suonio}, {Lagache}, {L{\"a}hteenm{\"a}ki}, {Lamarre},
  {Lange}, {Lasenby}, {Laureijs}, {Lawrence}, {Leach}, {Leahy}, {Leonardi},
  {Leroy}, {Lilje}, {Linden-V{\o}rnle}, {L{\'o}pez-Caniego}, {Lowe}, {Lubin},
  {Mac{\'\i}as-P{\'e}rez}, {Maciaszek}, {MacTavish}, {Maffei}, {Maino},
  {Mandolesi}, {Mann}, {Maris}, {Mart{\'\i}nez-Gonz{\'a}lez}, {Masi},
  {Massardi}, {Matarrese}, {Matthai}, {Mazzotta}, {McDonald}, {McGehee},
  {Meinhold}, {Melchiorri}, {Melin}, {Mendes}, {Mennella}, {Mevi},
  {Miniscalco}, {Mitra}, {Miville-Desch{\^e}nes}, {Moneti}, {Montier},
  {Morgante}, {Morisset}, {Mortlock}, {Munshi}, {Murphy}, {Naselsky}, {Natoli},
  {Netterfield}, {N{\o}rgaard-Nielsen}, {Noviello}, {Novikov}, {Novikov},
  {O'Dwyer}, {Ortiz}, {Osborne}, {Osuna}, {Oxborrow}, {Pajot}, {Paladini},
  {Partridge}, {Pasian}, {Passvogel}, {Patanchon}, {Pearson}, {Pearson},
  {Perdereau}, {Perotto}, {Perrotta}, {Piacentini}, {Piat}, {Pierpaoli},
  {Plaszczynski}, {Platania}, {Pointecouteau}, {Polenta}, {Ponthieu}, {Popa},
  {Poutanen}, {Pr{\'e}zeau}, {Prunet}, {Puget}, {Rachen}, {Reach}, {Rebolo},
  {Reinecke}, {Reix}, {Renault}, {Ricciardi}, {Riller}, {Ristorcelli}, {Rocha},
  {Rosset}, {Rowan-Robinson}, {Rubi{\~n}o-Mart{\'\i}n}, {Rusholme}, {Salerno},
  {Sandri}, {Santos}, {Savini}, {Schaefer}, {Scott}, {Seiffert}, {Shellard},
  {Simonetto}, {Smoot}, {Sozzi}, {Starck}, {Sternberg}, {Stivoli}, {Stolyarov},
  {Stompor}, {Stringhetti}, {Sudiwala}, {Sunyaev}, {Sygnet}, {Tapiador},
  {Tauber}, {Tavagnacco}, {Taylor}, {Terenzi}, {Texier}, {Toffolatti},
  {Tomasi}, {Torre}, {Tristram}, {Tuovinen}, {T{\"u}rler}, {Tuttlebee},
  {Umana}, {Valenziano}, {Valiviita}, {Varis}, {Vibert}, {Vielva}, {Villa},
  {Vittorio}, {Wade}, {Wandelt}, {Watson}, {White}, {White}, {Wilkinson},
  {Yvon}, {Zacchei}, \& {Zonca}}]{planck11}
{Planck Collaboration}, {Ade}, P.~A.~R., {Aghanim}, N., {et~al.} 2011, \aap,
  536, A1

\bibitem[{{Planck Collaboration} {et~al.}(2014){Planck Collaboration}, {Ade},
  {Aghanim}, {Armitage-Caplan}, {Arnaud}, {Ashdown}, {Atrio-Barand ela},
  {Aumont}, {Baccigalupi}, {Banday}, {Barreiro}, {Bartlett}, {Battaner},
  {Benabed}, {Beno{\^\i}t}, {Benoit-L{\'e}vy}, {Bernard}, {Bersanelli},
  {Bielewicz}, {Bobin}, {Bock}, {Bonaldi}, {Bond}, {Borrill}, {Bouchet},
  {Bridges}, {Bucher}, {Burigana}, {Butler}, {Calabrese}, {Cappellini},
  {Cardoso}, {Catalano}, {Challinor}, {Chamballu}, {Chary}, {Chen}, {Chiang},
  {Chiang}, {Christensen}, {Church}, {Clements}, {Colombi}, {Colombo},
  {Couchot}, {Coulais}, {Crill}, {Curto}, {Cuttaia}, {Danese}, {Davies},
  {Davis}, {de Bernardis}, {de Rosa}, {de Zotti}, {Delabrouille}, {Delouis},
  {D{\'e}sert}, {Dickinson}, {Diego}, {Dolag}, {Dole}, {Donzelli}, {Dor{\'e}},
  {Douspis}, {Dunkley}, {Dupac}, {Efstathiou}, {Elsner}, {En{\ss}lin},
  {Eriksen}, {Finelli}, {Forni}, {Frailis}, {Fraisse}, {Franceschi}, {Gaier},
  {Galeotta}, {Galli}, {Ganga}, {Giard}, {Giardino}, {Giraud-H{\'e}raud},
  {Gjerl{\o}w}, {Gonz{\'a}lez-Nuevo}, {G{\'o}rski}, {Gratton}, {Gregorio},
  {Gruppuso}, {Gudmundsson}, {Haissinski}, {Hamann}, {Hansen}, {Hanson},
  {Harrison}, {Henrot-Versill{\'e}}, {Hern{\'a}ndez-Monteagudo}, {Herranz},
  {Hildebrand t}, {Hivon}, {Hobson}, {Holmes}, {Hornstrup}, {Hou}, {Hovest},
  {Huffenberger}, {Jaffe}, {Jaffe}, {Jewell}, {Jones}, {Juvela},
  {Keih{\"a}nen}, {Keskitalo}, {Kisner}, {Kneissl}, {Knoche}, {Knox}, {Kunz},
  {Kurki-Suonio}, {Lagache}, {L{\"a}hteenm{\"a}ki}, {Lamarre}, {Lasenby},
  {Lattanzi}, {Laureijs}, {Lawrence}, {Leach}, {Leahy}, {Leonardi},
  {Le{\'o}n-Tavares}, {Lesgourgues}, {Lewis}, {Liguori}, {Lilje},
  {Linden-V{\o}rnle}, {L{\'o}pez-Caniego}, {Lubin}, {Mac{\'\i}as-P{\'e}rez},
  {Maffei}, {Maino}, {Mand olesi}, {Maris}, {Marshall}, {Martin},
  {Mart{\'\i}nez-Gonz{\'a}lez}, {Masi}, {Massardi}, {Matarrese}, {Matthai},
  {Mazzotta}, {Meinhold}, {Melchiorri}, {Melin}, {Mendes}, {Menegoni},
  {Mennella}, {Migliaccio}, {Millea}, {Mitra}, {Miville-Desch{\^e}nes},
  {Moneti}, {Montier}, {Morgante}, {Mortlock}, {Moss}, {Munshi}, {Murphy},
  {Naselsky}, {Nati}, {Natoli}, {Netterfield}, {N{\o}rgaard-Nielsen},
  {Noviello}, {Novikov}, {Novikov}, {O'Dwyer}, {Osborne}, {Oxborrow}, {Paci},
  {Pagano}, {Pajot}, {Paladini}, {Paoletti}, {Partridge}, {Pasian},
  {Patanchon}, {Pearson}, {Pearson}, {Peiris}, {Perdereau}, {Perotto},
  {Perrotta}, {Pettorino}, {Piacentini}, {Piat}, {Pierpaoli}, {Pietrobon},
  {Plaszczynski}, {Platania}, {Pointecouteau}, {Polenta}, {Ponthieu}, {Popa},
  {Poutanen}, {Pratt}, {Pr{\'e}zeau}, {Prunet}, {Puget}, {Rachen}, {Reach},
  {Rebolo}, {Reinecke}, {Remazeilles}, {Renault}, {Ricciardi}, {Riller},
  {Ristorcelli}, {Rocha}, {Rosset}, {Roudier}, {Rowan-Robinson},
  {Rubi{\~n}o-Mart{\'\i}n}, {Rusholme}, {Sandri}, {Santos}, {Savelainen},
  {Savini}, {Scott}, {Seiffert}, {Shellard}, {Spencer}, {Starck}, {Stolyarov},
  {Stompor}, {Sudiwala}, {Sunyaev}, {Sureau}, {Sutton}, {Suur-Uski}, {Sygnet},
  {Tauber}, {Tavagnacco}, {Terenzi}, {Toffolatti}, {Tomasi}, {Tristram},
  {Tucci}, {Tuovinen}, {T{\"u}rler}, {Umana}, {Valenziano}, {Valiviita}, {Van
  Tent}, {Vielva}, {Villa}, {Vittorio}, {Wade}, {Wandelt}, {Wehus}, {White},
  {White}, {Wilkinson}, {Yvon}, {Zacchei}, \& {Zonca}}]{planck14}
---. 2014, \aap, 571, A16

\bibitem[{{Planck Collaboration} {et~al.}(2016){Planck Collaboration}, {Ade},
  {Aghanim}, {Arnaud}, {Ashdown}, {Aumont}, {Baccigalupi}, {Banday},
  {Barreiro}, {Bartlett}, {Bartolo}, {Battaner}, {Battye}, {Benabed},
  {Beno{\^\i}t}, {Benoit-L{\'e}vy}, {Bernard}, {Bersanelli}, {Bielewicz},
  {Bock}, {Bonaldi}, {Bonavera}, {Bond}, {Borrill}, {Bouchet}, {Boulanger},
  {Bucher}, {Burigana}, {Butler}, {Calabrese}, {Cardoso}, {Catalano},
  {Challinor}, {Chamballu}, {Chary}, {Chiang}, {Chluba}, {Christensen},
  {Church}, {Clements}, {Colombi}, {Colombo}, {Combet}, {Coulais}, {Crill},
  {Curto}, {Cuttaia}, {Danese}, {Davies}, {Davis}, {de Bernardis}, {de Rosa},
  {de Zotti}, {Delabrouille}, {D{\'e}sert}, {Di Valentino}, {Dickinson},
  {Diego}, {Dolag}, {Dole}, {Donzelli}, {Dor{\'e}}, {Douspis}, {Ducout},
  {Dunkley}, {Dupac}, {Efstathiou}, {Elsner}, {En{\ss}lin}, {Eriksen},
  {Farhang}, {Fergusson}, {Finelli}, {Forni}, {Frailis}, {Fraisse},
  {Franceschi}, {Frejsel}, {Galeotta}, {Galli}, {Ganga}, {Gauthier}, {Gerbino},
  {Ghosh}, {Giard}, {Giraud-H{\'e}raud}, {Giusarma}, {Gjerl{\o}w},
  {Gonz{\'a}lez-Nuevo}, {G{\'o}rski}, {Gratton}, {Gregorio}, {Gruppuso},
  {Gudmundsson}, {Hamann}, {Hansen}, {Hanson}, {Harrison}, {Helou},
  {Henrot-Versill{\'e}}, {Hern{\'a}ndez-Monteagudo}, {Herranz}, {Hildebrand t},
  {Hivon}, {Hobson}, {Holmes}, {Hornstrup}, {Hovest}, {Huang}, {Huffenberger},
  {Hurier}, {Jaffe}, {Jaffe}, {Jones}, {Juvela}, {Keih{\"a}nen}, {Keskitalo},
  {Kisner}, {Kneissl}, {Knoche}, {Knox}, {Kunz}, {Kurki-Suonio}, {Lagache},
  {L{\"a}hteenm{\"a}ki}, {Lamarre}, {Lasenby}, {Lattanzi}, {Lawrence}, {Leahy},
  {Leonardi}, {Lesgourgues}, {Levrier}, {Lewis}, {Liguori}, {Lilje},
  {Linden-V{\o}rnle}, {L{\'o}pez-Caniego}, {Lubin}, {Mac{\'\i}as-P{\'e}rez},
  {Maggio}, {Maino}, {Mandolesi}, {Mangilli}, {Marchini}, {Maris}, {Martin},
  {Martinelli}, {Mart{\'\i}nez-Gonz{\'a}lez}, {Masi}, {Matarrese}, {McGehee},
  {Meinhold}, {Melchiorri}, {Melin}, {Mendes}, {Mennella}, {Migliaccio},
  {Millea}, {Mitra}, {Miville-Desch{\^e}nes}, {Moneti}, {Montier}, {Morgante},
  {Mortlock}, {Moss}, {Munshi}, {Murphy}, {Naselsky}, {Nati}, {Natoli},
  {Netterfield}, {N{\o}rgaard-Nielsen}, {Noviello}, {Novikov}, {Novikov},
  {Oxborrow}, {Paci}, {Pagano}, {Pajot}, {Paladini}, {Paoletti}, {Partridge},
  {Pasian}, {Patanchon}, {Pearson}, {Perdereau}, {Perotto}, {Perrotta},
  {Pettorino}, {Piacentini}, {Piat}, {Pierpaoli}, {Pietrobon}, {Plaszczynski},
  {Pointecouteau}, {Polenta}, {Popa}, {Pratt}, {Pr{\'e}zeau}, {Prunet},
  {Puget}, {Rachen}, {Reach}, {Rebolo}, {Reinecke}, {Remazeilles}, {Renault},
  {Renzi}, {Ristorcelli}, {Rocha}, {Rosset}, {Rossetti}, {Roudier},
  {Rouill{\'e} d'Orfeuil}, {Rowan-Robinson}, {Rubi{\~n}o-Mart{\'\i}n},
  {Rusholme}, {Said}, {Salvatelli}, {Salvati}, {Sandri}, {Santos},
  {Savelainen}, {Savini}, {Scott}, {Seiffert}, {Serra}, {Shellard}, {Spencer},
  {Spinelli}, {Stolyarov}, {Stompor}, {Sudiwala}, {Sunyaev}, {Sutton},
  {Suur-Uski}, {Sygnet}, {Tauber}, {Terenzi}, {Toffolatti}, {Tomasi},
  {Tristram}, {Trombetti}, {Tucci}, {Tuovinen}, {T{\"u}rler}, {Umana},
  {Valenziano}, {Valiviita}, {Van Tent}, {Vielva}, {Villa}, {Wade}, {Wandelt},
  {Wehus}, {White}, {White}, {Wilkinson}, {Yvon}, {Zacchei}, \&
  {Zonca}}]{planck16}
---. 2016, \aap, 594, A13

\bibitem[{{Planck Collaboration} {et~al.}(2020){Planck Collaboration},
  {Aghanim}, {Akrami}, {Ashdown}, {Aumont}, {Baccigalupi}, {Ballardini},
  {Banday}, {Barreiro}, {Bartolo}, {Basak}, {Battye}, {Benabed}, {Bernard},
  {Bersanelli}, {Bielewicz}, {Bock}, {Bond}, {Borrill}, {Bouchet}, {Boulanger},
  {Bucher}, {Burigana}, {Butler}, {Calabrese}, {Cardoso}, {Carron},
  {Challinor}, {Chiang}, {Chluba}, {Colombo}, {Combet}, {Contreras}, {Crill},
  {Cuttaia}, {de Bernardis}, {de Zotti}, {Delabrouille}, {Delouis}, {Di
  Valentino}, {Diego}, {Dor{\'e}}, {Douspis}, {Ducout}, {Dupac}, {Dusini},
  {Efstathiou}, {Elsner}, {En{\ss}lin}, {Eriksen}, {Fantaye}, {Farhang},
  {Fergusson}, {Fernandez-Cobos}, {Finelli}, {Forastieri}, {Frailis},
  {Fraisse}, {Franceschi}, {Frolov}, {Galeotta}, {Galli}, {Ganga},
  {G{\'e}nova-Santos}, {Gerbino}, {Ghosh}, {Gonz{\'a}lez-Nuevo}, {G{\'o}rski},
  {Gratton}, {Gruppuso}, {Gudmundsson}, {Hamann}, {Handley}, {Hansen},
  {Herranz}, {Hildebrandt}, {Hivon}, {Huang}, {Jaffe}, {Jones}, {Karakci},
  {Keih{\"a}nen}, {Keskitalo}, {Kiiveri}, {Kim}, {Kisner}, {Knox},
  {Krachmalnicoff}, {Kunz}, {Kurki-Suonio}, {Lagache}, {Lamarre}, {Lasenby},
  {Lattanzi}, {Lawrence}, {Le Jeune}, {Lemos}, {Lesgourgues}, {Levrier},
  {Lewis}, {Liguori}, {Lilje}, {Lilley}, {Lindholm}, {L{\'o}pez-Caniego},
  {Lubin}, {Ma}, {Mac{\'\i}as-P{\'e}rez}, {Maggio}, {Maino}, {Mandolesi},
  {Mangilli}, {Marcos-Caballero}, {Maris}, {Martin}, {Martinelli},
  {Mart{\'\i}nez-Gonz{\'a}lez}, {Matarrese}, {Mauri}, {McEwen}, {Meinhold},
  {Melchiorri}, {Mennella}, {Migliaccio}, {Millea}, {Mitra},
  {Miville-Desch{\^e}nes}, {Molinari}, {Montier}, {Morgante}, {Moss}, {Natoli},
  {N{\o}rgaard-Nielsen}, {Pagano}, {Paoletti}, {Partridge}, {Patanchon},
  {Peiris}, {Perrotta}, {Pettorino}, {Piacentini}, {Polastri}, {Polenta},
  {Puget}, {Rachen}, {Reinecke}, {Remazeilles}, {Renzi}, {Rocha}, {Rosset},
  {Roudier}, {Rubi{\~n}o-Mart{\'\i}n}, {Ruiz-Granados}, {Salvati}, {Sandri},
  {Savelainen}, {Scott}, {Shellard}, {Sirignano}, {Sirri}, {Spencer},
  {Sunyaev}, {Suur-Uski}, {Tauber}, {Tavagnacco}, {Tenti}, {Toffolatti},
  {Tomasi}, {Trombetti}, {Valenziano}, {Valiviita}, {Van Tent}, {Vibert},
  {Vielva}, {Villa}, {Vittorio}, {Wand elt}, {Wehus}, {White}, {White},
  {Zacchei}, \& {Zonca}}]{planck20}
{Planck Collaboration}, {Aghanim}, N., {Akrami}, Y., {et~al.} 2020, \aap, 641,
  A6

\bibitem[{{Price} \& {Bate}(2007)}]{price07}
{Price}, D.~J., \& {Bate}, M.~R. 2007, \mnras, 377, 77

\bibitem[{{Riaz} {et~al.}(2018){Riaz}, {Bovino}, {Vanaverbeke}, \&
  {Schleicher}}]{riaz18}
{Riaz}, R., {Bovino}, S., {Vanaverbeke}, S., \& {Schleicher}, D.~R.~G. 2018,
  \mnras, 479, 667

\bibitem[{{Riaz} {et~al.}(2022){Riaz}, {Hartwig}, \& {Latif}}]{riaz22}
{Riaz}, S., {Hartwig}, T., \& {Latif}, M.~A. 2022, arXiv e-prints,
  arXiv:2208.01673

\bibitem[{{Ripamonti}(2007)}]{ripamonti07hd}
{Ripamonti}, E. 2007, \mnras, 376, 709

\bibitem[{{Ripamonti} \& {Abel}(2004)}]{ra04}
{Ripamonti}, E., \& {Abel}, T. 2004, \mnras, 348, 1019

\bibitem[{{Rydberg} {et~al.}(2013){Rydberg}, {Zackrisson}, {Lundqvist}, \&
  {Scott}}]{rydberg13}
{Rydberg}, C.-E., {Zackrisson}, E., {Lundqvist}, P., \& {Scott}, P. 2013,
  \mnras, 429, 3658

\bibitem[{{Saigo} {et~al.}(2008){Saigo}, {Tomisaka}, \& {Matsumoto}}]{saigo08}
{Saigo}, K., {Tomisaka}, K., \& {Matsumoto}, T. 2008, \apj, 674, 997

\bibitem[{{Schauer} {et~al.}(2020){Schauer}, {Drory}, \& {Bromm}}]{sdb20}
{Schauer}, A. T.~P., {Drory}, N., \& {Bromm}, V. 2020, arXiv e-prints,
  arXiv:2007.02946

\bibitem[{{Schlaufman} {et~al.}(2018){Schlaufman}, {Thompson}, \&
  {Casey}}]{schlaufman18}
{Schlaufman}, K.~C., {Thompson}, I.~B., \& {Casey}, A.~R. 2018, \apj, 867, 98

\bibitem[{{Schleicher} {et~al.}(2009){Schleicher}, {Banerjee}, \&
  {Klessen}}]{schleicher09}
{Schleicher}, D. R.~G., {Banerjee}, R., \& {Klessen}, R.~S. 2009, \apj, 692,
  236

\bibitem[{{Schleicher} {et~al.}(2010){Schleicher}, {Banerjee}, {Sur},
  {Arshakian}, {Klessen}, {Beck}, \& {Spaans}}]{schleicher2010}
{Schleicher}, D.~R.~G., {Banerjee}, R., {Sur}, S., {et~al.} 2010, \aap, 522,
  A115

\bibitem[{{Schneider} {et~al.}(2012){Schneider}, {Omukai}, {Limongi},
  {Ferrara}, {Salvaterra}, {Chieffi}, \& {Bianchi}}]{schneider12}
{Schneider}, R., {Omukai}, K., {Limongi}, M., {et~al.} 2012, \mnras, 423, L60

\bibitem[{{Sharda} {et~al.}(2019){Sharda}, {Krumholz}, \&
  {Federrath}}]{sharda19}
{Sharda}, P., {Krumholz}, M.~R., \& {Federrath}, C. 2019, \mnras, 490, 513

\bibitem[{{Shu}(1977)}]{shu77}
{Shu}, F.~H. 1977, \apj, 214, 488

\bibitem[{{Springel} {et~al.}(2020){Springel}, {Pakmor}, {Zier}, \&
  {Reinecke}}]{springel20}
{Springel}, V., {Pakmor}, R., {Zier}, O., \& {Reinecke}, M. 2020, arXiv
  e-prints, arXiv:2010.03567

\bibitem[{{Stacy} \& {Bromm}(2014)}]{stacy14}
{Stacy}, A., \& {Bromm}, V. 2014, \apj, 785, 73

\bibitem[{{Stacy} {et~al.}(2010){Stacy}, {Greif}, \& {Bromm}}]{stacy10}
{Stacy}, A., {Greif}, T.~H., \& {Bromm}, V. 2010, \mnras, 403, 45

\bibitem[{{Stacy} {et~al.}(2013){Stacy}, {Greif}, {Klessen}, {Bromm}, \&
  {Loeb}}]{stacy13}
{Stacy}, A., {Greif}, T.~H., {Klessen}, R.~S., {Bromm}, V., \& {Loeb}, A. 2013,
  \mnras, 431, 1470

\bibitem[{{Stacy} {et~al.}(2022){Stacy}, {McKee}, {Lee}, {Klein}, \&
  {Li}}]{stacy22}
{Stacy}, A., {McKee}, C.~F., {Lee}, A.~T., {Klein}, R.~I., \& {Li}, P.~S. 2022,
  \mnras, 511, 5042

\bibitem[{{Sterzik} {et~al.}(2003){Sterzik}, {Durisen}, \& {Zinnecker}}]{sdz03}
{Sterzik}, M.~F., {Durisen}, R.~H., \& {Zinnecker}, H. 2003, \aap, 411, 91

\bibitem[{{Suda} {et~al.}(2021){Suda}, {Saitoh}, {Moritani}, {Matsuno}, \&
  {Shigeyama}}]{suda21}
{Suda}, T., {Saitoh}, T.~R., {Moritani}, Y., {Matsuno}, T., \& {Shigeyama}, T.
  2021, \pasj, 73, 609

\bibitem[{{Sugimura} {et~al.}(2020){Sugimura}, {Matsumoto}, {Hosokawa},
  {Hirano}, \& {Omukai}}]{sugimura20}
{Sugimura}, K., {Matsumoto}, T., {Hosokawa}, T., {Hirano}, S., \& {Omukai}, K.
  2020, \apjl, 892, L14

\bibitem[{{Sur} {et~al.}(2012){Sur}, {Federrath}, {Schleicher}, {Banerjee}, \&
  {Klessen}}]{sur12}
{Sur}, S., {Federrath}, C., {Schleicher}, D. R.~G., {Banerjee}, R., \&
  {Klessen}, R.~S. 2012, \mnras, 423, 3148

\bibitem[{{Sur} {et~al.}(2010){Sur}, {Schleicher}, {Banerjee}, {Federrath}, \&
  {Klessen}}]{sur10}
{Sur}, S., {Schleicher}, D.~R.~G., {Banerjee}, R., {Federrath}, C., \&
  {Klessen}, R.~S. 2010, \apjl, 721, L134

\bibitem[{{Susa}(2019)}]{susa19}
{Susa}, H. 2019, \apj, 877, 99

\bibitem[{{Susa} {et~al.}(1998){Susa}, {Uehara}, {Nishi}, \& {Yamada}}]{susa98}
{Susa}, H., {Uehara}, H., {Nishi}, R., \& {Yamada}, M. 1998, Progress of
  Theoretical Physics, 100, 63

\bibitem[{{Susa} {et~al.}(2009){Susa}, {Umemura}, \& {Hasegawa}}]{susa09}
{Susa}, H., {Umemura}, M., \& {Hasegawa}, K. 2009, \apj, 702, 480

\bibitem[{{Suto} \& {Silk}(1988)}]{suto88}
{Suto}, Y., \& {Silk}, J. 1988, \apj, 326, 527

\bibitem[{{Suzuki}(2018)}]{suzuki18}
{Suzuki}, T.~K. 2018, \pasj, 70, 34

\bibitem[{{Tanikawa} {et~al.}(2022){Tanikawa}, {Chiaki}, {Kinugawa}, {Suwa}, \&
  {Tominaga}}]{tanikawa22}
{Tanikawa}, A., {Chiaki}, G., {Kinugawa}, T., {Suwa}, Y., \& {Tominaga}, N.
  2022, \pasj, 74, 521

\bibitem[{{Tegmark} {et~al.}(2006){Tegmark}, {Eisenstein}, {Strauss},
  {Weinberg}, {Blanton}, {Frieman}, {Fukugita}, {Gunn}, {Hamilton}, {Knapp},
  {Nichol}, {Ostriker}, {Padmanabhan}, {Percival}, {Schlegel}, {Schneider},
  {Scoccimarro}, {Seljak}, {Seo}, {Swanson}, {Szalay}, {Vogeley}, {Yoo},
  {Zehavi}, {Abazajian}, {Anderson}, {Annis}, {Bahcall}, {Bassett}, {Berlind},
  {Brinkmann}, {Budavari}, {Castander}, {Connolly}, {Csabai}, {Doi},
  {Finkbeiner}, {Gillespie}, {Glazebrook}, {Hennessy}, {Hogg}, {Ivezi{\'c}},
  {Jain}, {Johnston}, {Kent}, {Lamb}, {Lee}, {Lin}, {Loveday}, {Lupton},
  {Munn}, {Pan}, {Park}, {Peoples}, {Pier}, {Pope}, {Richmond}, {Rockosi},
  {Scranton}, {Sheth}, {Stebbins}, {Stoughton}, {Szapudi}, {Tucker}, {vand en
  Berk}, {Yanny}, \& {York}}]{tegmark06}
{Tegmark}, M., {Eisenstein}, D.~J., {Strauss}, M.~A., {et~al.} 2006, \prd, 74,
  123507

\bibitem[{{Tokuoka} {et~al.}(2022){Tokuoka}, {Inoue}, {Hashimoto}, {Ellis},
  {Laporte}, {Sugahara}, {Matsuo}, {Tamura}, {Fudamoto}, {Moriwaki},
  {Roberts-Borsani}, {Shimizu}, {Yamanaka}, {Yoshida}, {Zackrisson}, \&
  {Zheng}}]{tokuoka22}
{Tokuoka}, T., {Inoue}, A.~K., {Hashimoto}, T., {et~al.} 2022, \apjl, 933, L19

\bibitem[{{Truelove} {et~al.} (1998)}]{Truelove98} {Truelove} J K, {Klein} R I, {McKee} C. F., {Holliman} J. H., {Howell} L. H., {Greenough} J. A. \& {Woods} D. T. 1998, \apj,  495, 821

\bibitem[{{Tumlinson}(2010)}]{tumlinson10}
{Tumlinson}, J. 2010, \apj, 708, 1398

\bibitem[{{Turk} {et~al.}(2009){Turk}, {Abel}, \& {O'Shea}}]{tao09}
{Turk}, M.~J., {Abel}, T., \& {O'Shea}, B. 2009, Science, 325, 601

\bibitem[{{Turk} {et~al.}(2012){Turk}, {Oishi}, {Abel}, \& {Bryan}}]{turk12}
{Turk}, M.~J., {Oishi}, J.~S., {Abel}, T., \& {Bryan}, G.~L. 2012, \apj, 745,
  154

\bibitem[{{Umeda} {et~al.}(2016){Umeda}, {Hosokawa}, {Omukai}, \&
  {Yoshida}}]{umeda16}
{Umeda}, H., {Hosokawa}, T., {Omukai}, K., \& {Yoshida}, N. 2016, \apjl, 830,
  L34

\bibitem[{{Wang} {et~al.}(2020){Wang}, {Bose}, {Frenk}, {Gao}, {Jenkins},
  {Springel}, \& {White}}]{wang20}
{Wang}, J., {Bose}, S., {Frenk}, C.~S., {et~al.} 2020, \nat, 585, 39

\bibitem[{{Welch} {et~al.}(2022){Welch}, {Coe}, {Diego}, {Zitrin},
  {Zackrisson}, {Dimauro}, {Jim{\'e}nez-Teja}, {Kelly}, {Mahler}, {Oguri},
  {Timmes}, {Windhorst}, {Florian}, {de Mink}, {Avila}, {Anderson}, {Bradley},
  {Sharon}, {Vikaeus}, {McCandliss}, {Brada{\v{c}}}, {Rigby}, {Frye}, {Toft},
  {Strait}, {Trenti}, {Sharma}, {Andrade-Santos}, \& {Broadhurst}}]{welch22}
{Welch}, B., {Coe}, D., {Diego}, J.~M., {et~al.} 2022, \nat, 603, 815

\bibitem[{{Wells} \& {Norman}(2021)}]{wells21}
{Wells}, A.~I., \& {Norman}, M.~L. 2021, arXiv e-prints, arXiv:2111.10651

\bibitem[{{Welsh} {et~al.}(2019){Welsh}, {Cooke}, \& {Fumagalli}}]{welsh19}
{Welsh}, L., {Cooke}, R., \& {Fumagalli}, M. 2019, \mnras, 487, 3363

\bibitem[{{Whalen} {et~al.}(2004){Whalen}, {Abel}, \& {Norman}}]{whalen04}
{Whalen}, D., {Abel}, T., \& {Norman}, M.~L. 2004, \apj, 610, 14

\bibitem[{{Whalen} {et~al.}(2008){Whalen}, {O'Shea}, {Smidt}, \&
  {Norman}}]{whalen08rad}
{Whalen}, D., {O'Shea}, B.~W., {Smidt}, J., \& {Norman}, M.~L. 2008, \apj, 679,
  925

\bibitem[{{Whalen} {et~al.}(2014){Whalen}, {Smidt}, {Even}, {Woosley}, {Heger},
  {Stiavelli}, \& {Fryer}}]{whalen14}
{Whalen}, D.~J., {Smidt}, J., {Even}, W., {et~al.} 2014, \apj, 781, 106

\bibitem[{{White} \& {Springel}(2000)}]{ws00}
{White}, S. D.~M., \& {Springel}, V. 2000, in The First Stars, ed. A.~{Weiss},
  T.~G. {Abel}, \& V.~{Hill}, 327

\bibitem[{{Wise} \& {Abel}(2008)}]{wise08}
{Wise}, J.~H., \& {Abel}, T. 2008, \apj, 685, 40

\bibitem[{{Wise} {et~al.}(2012){Wise}, {Abel}, {Turk}, {Norman}, \&
  {Smith}}]{wise12}
{Wise}, J.~H., {Abel}, T., {Turk}, M.~J., {Norman}, M.~L., \& {Smith}, B.~D.
  2012, \mnras, 427, 311

\bibitem[{{Wollenberg} {et~al.}(2020){Wollenberg}, {Glover}, {Clark}, \&
  {Klessen}}]{wollenberg20}
{Wollenberg}, K. M.~J., {Glover}, S. C.~O., {Clark}, P.~C., \& {Klessen}, R.~S.
  2020, \mnras, 494, 1871

\bibitem[{{Woods} {et~al.}(2017){Woods}, {Heger}, {Whalen}, {Haemmerl{\'e}}, \&
  {Klessen}}]{woods17}
{Woods}, T.~E., {Heger}, A., {Whalen}, D.~J., {Haemmerl{\'e}}, L., \&
  {Klessen}, R.~S. 2017, \apjl, 842, L6

\bibitem[{{Xu} \& {Stone}(2019)}]{xu19}
{Xu}, W., \& {Stone}, J.~M. 2019, \mnras, 488, 5162

\bibitem[{{Yoshida} {et~al.}(2006){Yoshida}, {Omukai}, {Hernquist}, \&
  {Abel}}]{yoha06}
{Yoshida}, N., {Omukai}, K., {Hernquist}, L., \& {Abel}, T. 2006, \apj, 652, 6








\end{thebibliography}

\end{document}